\documentclass[aps,prb,groupedaddress,
onecolumn]{revtex4-2}
\usepackage{graphicx,bm,wasysym,color,ifsym,amssymb}

\def\be{\begin{equation}}  
\def\ee{\end{equation}}  
\def\ba{\begin{eqnarray}}  
\def\ea{\end{eqnarray}}  
\def\bc{\begin{center}}  
\def\ec{\end{center}}  
\def\p{\partial}

\begin{document}

\title{Nonperturbative quasiclassical theory of graphene photoconductivity}

\author{S. A. Mikhailov}
\email[Email: ]{sergey.mikhailov@physik.uni-augsburg.de}
\affiliation{Institute of Physics, University of Augsburg, D-86135 Augsburg, Germany}

\date{\today}

\begin{abstract}
We present a nonperturbative quasi-classical theory of graphene photoconductivity. We consider the influence of low-frequency (microwave, terahertz, mid-infrared) radiation on the static conductivity of a uniform graphene layer and calculate its photoconductivity as a function of frequency, polarization and strength of the external ac electric field, as well as on the material properties (electron density, scattering time) and temperature. The theory is valid at frequencies $\hbar\omega\lesssim 2E_F$ and at arbitrarily strong ac electric fields. We compare our results with those of the third-order perturbation theory and determine the applicability range of the perturbative solutions.
\end{abstract}


\maketitle

\tableofcontents

\section{Introduction\label{sec:intro}}

A distinctive feature of graphene \cite{Neto09} is the linear energy dispersion of its electrons and holes \cite{Wallace47}. It was predicted \cite{Mikhailov07e} that this feature should lead to a strongly nonlinear electrodynamic and optical response of this material in relatively weak external 
electric fields. Subsequent experimental and theoretical studies confirmed this prediction, see, e.g. Refs. \cite{Mikhailov08a,Hendry10,Bykov12,Mikhailov11c,Cheng14a,Cheng14b,Cheng15,Mikhailov16a,Wang16,Mikhailov17a,Savostianova17a,Cheng17,Alexander17,MariniAbajo17,Savostianova18a,Mikhailov19b}. Currently, nonlinear electrodynamics and optics of graphene is a hot and quickly developing area of research.

Theoretically, the nonlinear electrodynamic response of graphene was mainly considered within the frameworks of the perturbation theory \cite{Mikhailov07e,Mikhailov08a,Cheng14a,Cheng14b,Cheng15,Mikhailov16a,Wang16,Cheng17}. Within such a theory, the electric current $\bm j(t)$ is expanded in a Taylor series up to the third order in powers of the electric field $\bm E(t)$,
\be
j_{\alpha}(t)=
\int_{-\infty}^\infty d\omega_1 
\sigma_{\alpha\beta}^{(1)}(\omega_1)
E^\beta_{\omega_1} 
e^{-i\omega_1  t} +
\int_{-\infty}^\infty d\omega_1 
\int_{-\infty}^\infty d\omega_2  
\int_{-\infty}^\infty d\omega_3 
\sigma_{\alpha\beta\gamma\delta}^{(3)}(\omega_1,\omega_2,\omega_3)
E^\beta_{\omega_1} 
E^\gamma_{\omega_2} 
E^\delta_{\omega_3} 
e^{-i(\omega_1 +\omega_2+\omega_3) t}+\dots 
\label{Taylor}
\ee
where $E^\beta_{\omega}$ are Fourier components of the time-dependent electric field 
\be 
E_{\alpha}(t)=\int_{-\infty}^\infty d\omega  E_\omega^{\alpha} e^{-i\omega t}.
\label{fieldFourier}
\ee 
The linear and third-order conductivities $\sigma_{\alpha\beta}^{(1)}(\omega_1)$ and $\sigma_{\alpha\beta\gamma\delta}^{(3)}(\omega_1,\omega_2,\omega_3)$ have been calculated as functions of frequencies $\omega_1$, $\omega_2$, $\omega_3$, Fermi energy $E_F$ and scattering parameters in Refs. \cite{Gusynin07b,Falkovsky07a,Mikhailov07d} and \cite{Cheng14a,Cheng15,Mikhailov16a} respectively. The latter describes a large number of different physical effects, such as the third harmonic generation (at $\omega_1=\omega_2=\omega_3$), saturable absorption and Kerr effect (at $\omega_1=\omega_2=-\omega_3$), direct current induced second harmonic generation (at $\omega_1=\omega_2$ and $\omega_3=0$), static photoconductivity (at $\omega_1=-\omega_2$ and $\omega_3=0$) and many other.

The perturbation approach allows to obtain corrections to results of the linear theory, but its applicability is also restricted: the Taylor expansion (\ref{Taylor}) implies that the third order term is smaller than the first one. However, in many experiments the external electric field is so strong that the third-order theory becomes insufficient for a proper description of the nonlinear response of the material. In graphene this may happen already in electric fields of order of $1-3$ kV/cm \cite{Mikhailov07e,Mikhailov08a}. In such cases a nonperturbative theory is required.

In Ref. \cite{Mikhailov17a} we have developed a \textit{nonperturbative quasiclassical} theory of the nonlinear electrodynamic response of uniform graphene. The kinetic Boltzmann equation was solved there in the relaxation time approximation, which allowed to describe the graphene response to arbitrarily strong external electric fields at ``low'' (microwave, terahertz, infrared) frequencies satisfying the condition $\hbar\omega\lesssim 2E_F$. In Ref. \cite{Mikhailov17a} we applied our general results to the case, when a strong ac electric field $E_\omega\sin\omega t$ acts on the system, and analyzed the odd harmonics generation and Kerr effects. 
 
In this paper we apply the theory \cite{Mikhailov17a} to the analysis of another physical effect, the static photoconductivity of graphene. Without irradiation, the graphene response to a weak external dc electric field $\bm E_0$ is described by the conventional isotropic Drude conductivity $\sigma_0$, $\bm j_0=\sigma_0\bm E_0$. Now we assume that, in addition to the weak dc field $\bm E_0$, a strong monochromatic ac electric field $\bm E_{\rm ac}(t)$ acts on graphene electrons,
\be 
\bm E(t)=\bm E_0+\bm E_{\rm ac}(t),
\label{ext-field}
\ee 
and calculate the resulting time-averaged direct current in the linear order in $\bm E_0$, 
\be 
j_\alpha^0=\sigma_{\alpha\beta}^{\mathrm{ph}}(\bm E_{\rm ac}) E_\beta^0.
\label{ph-cond}
\ee
The photoconductivity tensor $\sigma_{\alpha\beta}^{\mathrm{ph}}(\bm E_{\rm ac})$ here is a function of graphene parameters as well as the amplitude, frequency and polarization of the incident radiation. We analyze these dependencies and compare the nonperturbative results with those obtained within the third-order perturbation theory. 

Theoretically the photoconductivity of graphene was studied in a number of publications, see, e.g., Refs. \cite{Vasko08,Romanets10,Bao10,Trushin11,Shao15,Singh18,Ryzhii19}. Most of these papers mainly focused on the photoresponse of \textit{intrinsic} graphene ($E_F\approx 0$) generated by the \textit{interband} excitation of charge carriers, $\hbar\omega\gtrsim 2E_F$, which is typically relevant for near-IR/optical excitation. Here we concentrate on the opposite limit $\hbar\omega\lesssim 2E_F$ where the interband transitions can be ignored and which is relevant for microwave/terahertz/mid-IR excitation of the system. For example, if the density of graphene electrons is $\sim 10^{13}$ cm$^{-2}$, the condition $\hbar\omega\lesssim 2E_F$ is satisfied at frequencies $f\lesssim 175$ THz or the wavelength $\lambda\gtrsim 1.7$ $\mu$m. 

\section{Theory and results\label{sec:Theory}}

\subsection{General formulas\label{sec:GenFormulas}}

Within the quasiclassical approach the nonlinear electrodynamic response of graphene to the field (\ref{ext-field}) can be described by the Boltzmann equation in the relaxation time approximation,
\be 
\frac{\p f(\bm p,t)}{\p t}-e\bm E(t)\frac{\p f(\bm p,t)}{\p \bm p}=-\frac{f(\bm p,t)-f_0(\bm p)}{\tau}\label{be},
\ee
where 
\be 
f_0({\bm p})=\left[1+\exp\left(\frac {E_{\bm p}-\mu}T\right)\right]^{-1}
\ee
is the Fermi-Dirac distribution function, and $\tau$ is the momentum relaxation time, which we assume to be energy independent. For definiteness we will consider graphene electrons assuming that the chemical potential is positive, $\mu>0$, and will describe their spectrum near Dirac points by the linear energy dispersion 
\be 
E_{\bm p}=v_F|\bm p|=v_F\sqrt{p_x^2+p_y^2},
\label{spectrum}
\ee
with $v_F\approx 10^8$ cm/s being the  Fermi velocity. 

The solution of Eq. (\ref{be}) at arbitrary electric fields $\bm E(t)$ has the form \cite{Ignatov76,Mikhailov17a} 
\be 
f(\bm p,t)=
 \int^{\infty}_0 e^{-\xi} f_0\big(\bm p- \bm p_0(t,\xi)\big)   d \xi,
\label{solution}
\ee
where the vector
\be 
\bm p_0(t,\xi)=-e\int_{t-\xi\tau}^t  {\bm E}(t') dt'
\label{p0}
\ee
is determined by the external electric field. The induced electric current is then found by summation over occupied quantum states 
\be 
\bm j(t)=-\frac{e}{S}\sum_{\bm p\sigma v} \frac{\p E_{\bm p}}{\p \bm p} f(\bm p,t)
=
-\frac{e}{S}\sum_{\bm p\sigma v} \frac{\p E_{\bm p}}{\p \bm p} \int^{\infty}_0 e^{-\xi} f_0\big(\bm p- \bm p_0(t,\xi)\big)   d \xi,\label{current}
\ee
where $S$ is the sample area, $\sigma$ and $v$ are the spin and valley quantum numbers. Substituting (\ref{p0}) and (\ref{spectrum}) into Eq. (\ref{current}) we get, after some algebra, the following expression for the electric current (a similar calculation can be found in \cite{Mikhailov17a})
\ba 
\bm j(t)=
\frac{eg_sg_v\pi }{(2\pi\hbar)^2v_F}
\frac 1{4T}\int_{-\infty}^\infty \frac{E_F^2dE_F}{\cosh^2\left(\frac{\mu-E_F}{2T}\right)} 
\int^{\infty}_0 e^{-\xi}  d \xi 
\bm P(t,\xi,E_F)N\big[P(t,\xi,E_F)\big].
\label{genresforj}
\ea
In Eq. (\ref{genresforj}), $g_s$ and $g_v$ are the spin and valley degeneracies, $g_s=g_v=2$,
\be 
\bm P(t,\xi,E_F)=- \frac{\bm p_{0}(t,\xi)}{p_F},\ \ P(t,\xi,E_F)=|\bm P(t,\xi,E_F)|,
\label{vectorP}
\ee
$p_F=E_F/v_F$ is the Fermi momentum, and the function $N(x)$ is defined as
\be 
N(x)=
\frac{1}{ \sqrt{1+x^2}}\ {_2F_1}\left(\frac 14,\frac 34;2; \left(\frac{2x}{1+x^2}\right)^2\right),
\ee
where ${_2F_1}(a,b;c;z)$ is the hypergeometric function. Equation (\ref{genresforj}) gives a general expression for the current, as a function of the chemical potential, temperature, and the scattering parameter $\tau$, at different time dependencies and polarizations of the external electric field. 

Before moving further let us discuss the temperature dependence of the current (\ref{genresforj}). At zero temperature $T=0$ the factor with the $\cosh$ function is reduced to the delta-function,
\be 
\lim_{T\to 0}\frac{1}{4T\cosh^2\left(\frac{\mu-E_F}{2T}\right)} =\delta(\mu-E_F).
\label{limit-delta}
\ee
At higher temperatures the current varies with $T$, but these changes are not very large. Indeed, in the quasiclassical theory the chemical potential should be considered to be large, to satisfy the condition $\hbar\omega\lesssim 2E_F$. For example, if $E_F$ is about $\sim 0.2$ eV or larger (this corresponds to electron densities larger than $\sim 3\times 10^{12}$ cm$^{-2}$), the condition $T\ll E_F$ is satisfied not only at the room temperature $T_0$ but also at $T$ exceeding $T_0$ by a factor $2-3$. Therefore, we can get accurate results assuming $T\ll E_F$ and using the following simplified expression for the current
\ba 
\bm j(t)=
\frac{eg_sg_v\pi E_F^2}{(2\pi\hbar)^2v_F}
\int^{\infty}_0 e^{-\xi}  d \xi 
\bm P(t,\xi,E_F)N\big[P(t,\xi,E_F)\big].
\label{genresforjT0}
\ea
Here we have used the limit (\ref{limit-delta}) and replaced $\mu$ by a more convenient designation $E_F$. In the rest of the paper except Section \ref{sec:temperature} we will use the simplified expression (\ref{genresforjT0}). If Section \ref{sec:temperature} we analyze the finite temperature effects using a more general formula (\ref{genresforj}) and show that temperature does not substantially influence the $T=0$ results indeed. 

Now we discuss results for the photoconductivity of graphene obtained from the general equation (\ref{genresforjT0}) in different limiting cases. We assume that the weak dc field is parallel to the $x$-axis, $\bm E_0\parallel\bm e_x$, and consider several possible polarizations of the strong ac electric field. 

\subsection{Linearly polarized light: Photoconductivity versus polarization angle\label{sec:LinearPolariz}}

First, let us consider the case when the ac field is linearly polarized, and the polarization plane of the incident radiation constitutes an angle $\theta$ with respect to the $\bm E_0$ field. Then we have  
\be 
\bm E(t)=\bm e_x (E_0+E_\omega\cos\theta\sin\omega t)+\bm e_y E_\omega\sin\theta\sin\omega t.
\ee
According to the definition (\ref{vectorP}), 
\be 
\bm P(t,\xi)=\bm e_x \left({\cal F}_0\xi+{\cal F}_\omega\cos\theta\frac{\sin(\omega\tau\xi/2)}{\omega\tau/2}\sin(\omega t-\omega\tau\xi/2)\right)
+\bm e_y {\cal F}_\omega\sin\theta\frac{\sin(\omega\tau\xi/2)}{\omega \tau/2}\sin(\omega t-\omega\tau\xi/2),
\label{Ppolarization}
\ee 
where we have introduced dimensionless quantities
\be
{\cal F}_0=\frac{e E_0\tau}{p_F},\ \  {\cal F}_\omega=\frac{eE_\omega\tau}{p_F},
\ee
characterizing the electric fields strength: the conditions ${\cal F}_{0,\omega}\ll 1$ and ${\cal F}_{0,\omega}\gtrsim 1$ correspond to the linear-response and nonlinear regimes, respectively \cite{Mikhailov07e}. Substituting (\ref{Ppolarization}) into (\ref{genresforjT0}) and averaging the resulting expression over time we obtain the absolute value of the direct current 
\be 
\left(\begin{array}{c}
j_x^0 \\ j_y^0 \\
\end{array}\right)=en_sv_F
\int^{\infty}_0 e^{-\xi}  d \xi 
\frac 1{2\pi}\int_{-\pi}^\pi dx 
\left(\begin{array}{c}
{\cal F}_0\xi+Z\cos\theta\sin x \\ Z\sin\theta\sin x \\
\end{array}\right)
N\left(\sqrt{\left({\cal F}_0\xi+Z\cos\theta\sin x\right)^2
+ \left(Z\sin\theta\sin x\right)^2}\right),
\label{current-arbitrarytheta}
\ee
where 
\be 
n_s=\frac{g_sg_vE_F^2}{4\pi\hbar^2v_F^2}
\ee
is the density of electrons in graphene, and we have introduced a short notation
\be 
Z\equiv Z({\cal F}_\omega,\omega\tau,\xi)={\cal F}_\omega\frac{\sin(\omega\tau\xi/2)}{\omega\tau/2}.\label{Z}
\ee
As seen from Eq. (\ref{current-arbitrarytheta}) the current flows both in $x$- and $y$-directions. In order to get compact expressions for components of the tensor $\sigma_{\alpha\beta}^{\rm ph}(\bm E_{\rm ac})$ it is convenient to introduce two functions 
\be 
{\cal A}({\cal F}_\omega,\omega\tau)=\int^{\infty}_0 \xi e^{-\xi}  d \xi 
\frac 2{\pi}\int_{0}^{\pi/2} dx 
N\left(Z\sin x\right),
\label{A}
\ee
\be 
{\cal B}({\cal F}_\omega,\omega\tau)=\int^{\infty}_0 \xi e^{-\xi}  d \xi 
\frac 2{\pi}\int_{0}^{\pi/2} dx \left(Z\sin x\right)^2
M\left(Z\sin x\right),
\label{B}
\ee
where the function $M(x)$ is related to the derivative of $N(x)$, $N'(x)=-xM(x)$, and is determined by the formula
\be 
M(x)=\frac{1}{ (1+x^2)^{3/2}}\left[ {_2F_1}\left(\frac 14,\frac 34;2; \left(\frac{2x}{1+x^2}\right)^2\right)-\frac{3}{4}
\frac{1-x^2}{(1+x^2)^{2} } \ {_2F_1}\left(\frac 54,\frac 74;3; \left(\frac{2x}{1+x^2}\right)^2\right)\right].
\label{funcM}
\ee
Taking the linear-response limit $E_0\to 0$ we get the components of the tensor $\sigma_{\alpha\beta}^{\rm ph}(\bm E_{\rm ac})$:
\be 
\frac{\sigma_{xx}^{\rm ph}({\cal F}_\omega,\omega\tau,\theta)}{\sigma_0} =
{\cal A}({\cal F}_\omega,\omega\tau)- \cos^2\theta
{\cal B}({\cal F}_\omega,\omega\tau),
\label{Pcond-ThetaXX}
\ee
and 
\be 
\frac{\sigma_{yx}^{\rm ph}({\cal F}_\omega,\omega\tau,\theta)}{\sigma_0} =
- \sin\theta\cos\theta 
{\cal B}({\cal F}_\omega,\omega\tau).
\label{Pcond-ThetaYX}
\ee

\begin{figure}[ht]
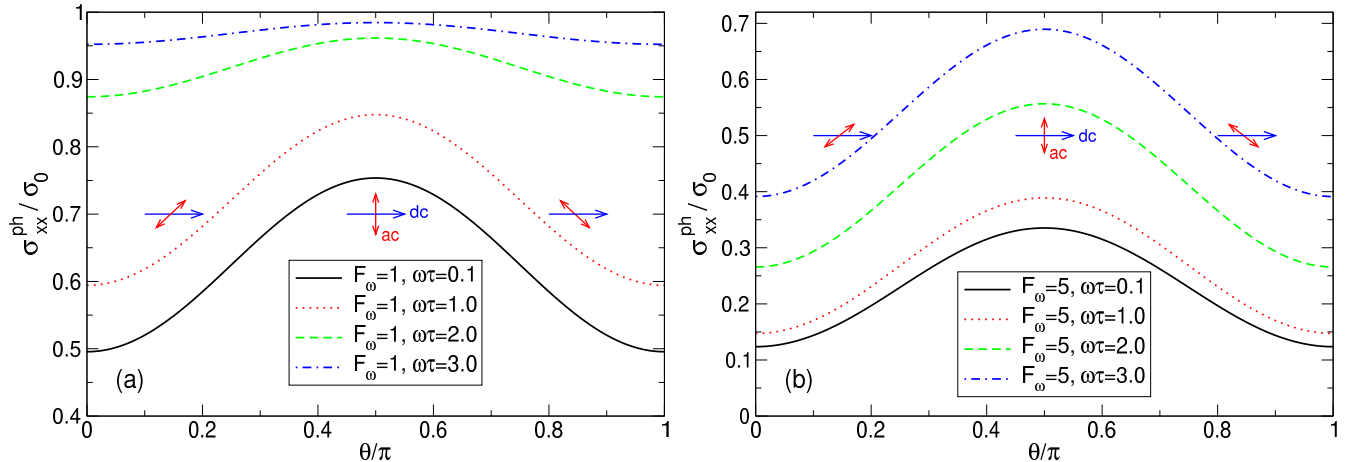

\includegraphics[width=0.49\textwidth]{SigmaXX-Theta_Fw1.eps}  
\includegraphics[width=0.49\textwidth]{SigmaXX-Theta_Fw5.eps}  
\caption{\label{fig:PCvsThetaXX} The photoconductivity (\ref{Pcond-ThetaXX}) as a function of $\theta$ at several values of $\omega\tau$ and at (a) the electric field parameter ${\cal F}_\omega=1$ and (b) ${\cal F}_\omega=5$. Blue and red arrows illustrate the mutual orientation of the dc and ac electric fields at different points of the $\theta$-axis.}
\end{figure}

Figure \ref{fig:PCvsThetaXX} illustrates the $\theta$-dependence of the diagonal photoconductivity $\sigma_{xx}^{\rm ph}({\cal F}_\omega,\omega\tau,\theta)$ at a few values of the electric field strength parameter ${\cal F}_\omega$ and the frequency parameter $\omega\tau$. First, one sees that $\sigma_{xx}^{\rm ph}$ is smaller than $\sigma_0$, i.e., the infinite uniform graphene layer is characterized by the \textit{negative} diagonal photoconductivity. Second, the influence of radiation on the conductive properties of the material can be very large: at a quite moderate value of the electric field parameter ${\cal F}_\omega\simeq 1$ the conductivity at low frequencies $\omega\tau\ll 1$ can be reduced by a factor of two, Figure \ref{fig:PCvsThetaXX}(a), black curve. At larger values of ${\cal F}_\omega$ the effect of radiation increases further: at ${\cal F}_\omega=5$ the conductivity changes by 80-90 \%, Figure \ref{fig:PCvsThetaXX}(b). Also, the effect is highly frequency dependent: it is highest at low frequencies  $\omega\tau\lesssim 1$ and decreases at $\omega\tau\gg 1$. The maximal reduction of $\sigma_{xx}^{\rm ph}$ is seen when the dc and ac electric fields are parallel to each other, at $\theta=0$ or $\pi$. At $\theta=\pi/2$ the conductivity change $\delta\sigma_{xx}^{\rm ph}$ is weaker, and the difference between $\delta\sigma_{xx}^{\rm ph}(\theta=0)$ and $\delta\sigma_{xx}^{\rm ph}(\theta=\pi/2)$ is comparable with the value of $\delta\sigma_{xx}^{\rm ph}(\theta=\pi/2)$ itself, i.e., the effect is quite sensitive to the polarization of the wave. 

The negative sign of the \textit{intraband} photoconductivity $\sigma_{xx}^{\rm ph}({\cal F}_\omega,\omega\tau,\theta)$ is explained by the linear energy dispersion of graphene electrons. Under intense irradiation electrons get additional energy $E$ from the ac electric field occupying quantum states with $E>E_F$, see Appendix \ref{app:distrfun}. As a result, the ``effective mass'' of electrons $\sim E/v_F^2$ increases and the intraband (Drude) conductivity decreases. 

\begin{figure}[ht]
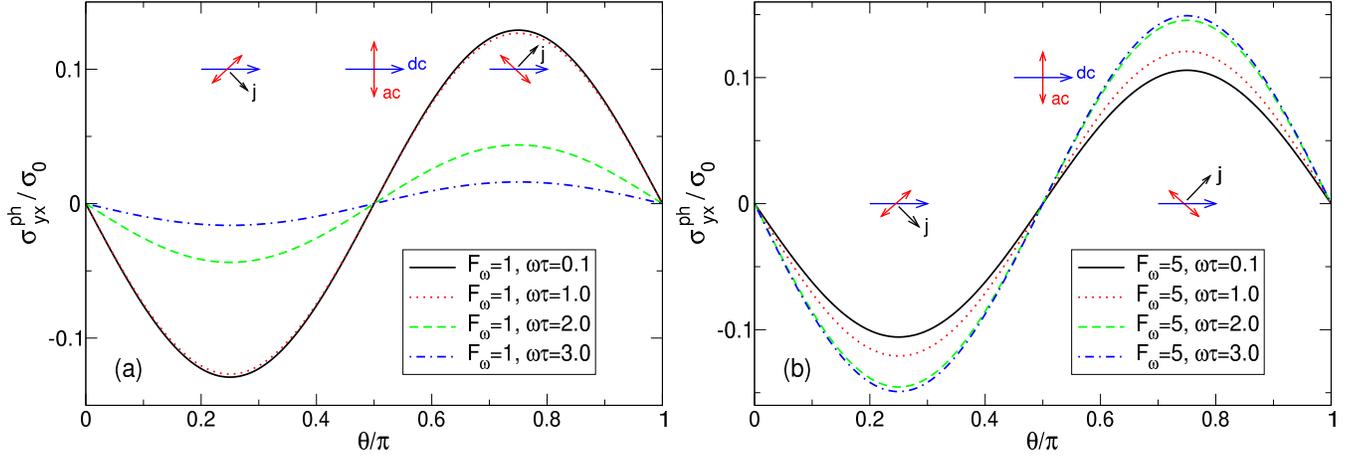

\includegraphics[width=0.49\textwidth]{SigmaYX-Theta_Fw1.eps}  
\includegraphics[width=0.49\textwidth]{SigmaYX-Theta_Fw5.eps}  
\caption{\label{fig:PCvsThetaYX} The photoconductivity (\ref{Pcond-ThetaYX}) as a function of $\theta$ at several values of $\omega\tau$ and at (a) the electric field parameter ${\cal F}_\omega=1$ and (b) ${\cal F}_\omega=5$. Blue and red arrows illustrate the mutual orientation of the dc and ac electric fields at different points of the $\theta$-axis. Black arrows show the direction of the wave induced photocurrent.}
\end{figure}

Figure \ref{fig:PCvsThetaYX} shows the $\theta$-dependence of the off-diagonal photoconductivity $\sigma_{yx}^{\rm ph}({\cal F}_\omega,\omega\tau,\theta)$ at the same values of ${\cal F}_\omega$ and $\omega\tau$. Now, if the direction of the ac field is parallel or perpendicular to the direction of the dc field ($\theta=0$, $\pi$ or $\pi/2$), the $y$-component of the photocurrent $j_y^0$ vanishes. If $\theta$ lies between 0 and $\pi/2$, the current $j_y^0$ is negative, while if $\pi/2<\theta<\pi$, it is positive; see the directions of the resulting photocurrent in Figure \ref{fig:PCvsThetaYX} (black arrows). The dependence of the transverse photoconductivity $\sigma_{yx}^{\rm ph}$ on ${\cal F}_\omega$ and $\omega\tau$ is less trivial and more interesting than that of $\sigma_{xx}^{\rm ph}$. First, one sees that the maximum ($\theta=3\pi/4$) low-frequency ($\omega\tau=0.1$) value of the transverse photoconductivity in moderate ac field ${\cal F}_\omega=1$ is \textit{larger} than in the strong field ${\cal F}_\omega=5$, $\sigma_{yx}^{\rm ph}({\cal F}_\omega=1)\approx 0.13$ vs. $\sigma_{yx}^{\rm ph}({\cal F}_\omega=5)\approx 0.107$, compare Figs. \ref{fig:PCvsThetaYX}(a) and \ref{fig:PCvsThetaYX}(b). Furthermore, the moderate-field value $\sigma_{yx}^{\rm ph}({\cal F}_\omega=1)$ decreases, while the high-field value $\sigma_{yx}^{\rm ph}({\cal F}_\omega=5)$ increases with the growing frequency: for example, $\sigma_{yx}^{\rm ph}({\cal F}_\omega=1,\omega\tau=3)\approx 0.016$ while $\sigma_{yx}^{\rm ph}({\cal F}_\omega=5,\omega\tau=3)\approx 0.15$. We investigate these interesting ${\cal F}_\omega$ and $\omega\tau$ dependencies further in Section \ref{sec:E&w-dependence} below.

\subsection{Elliptically polarized light: Photoconductivity versus ellipticity\label{sec:ElliptPolariz}}

Now, let us consider the case when the ac field is elliptically polarized, with the polarization ellipse axes parallel to the $x$- and $y$-directions. The dc field $\bm E_0$ is assumed to be parallel to the $x$ axis as before. Then we write the electric field in the form  
\be 
\bm E(t)=\bm e_x (E_0+E_\omega\cos\delta\cos\omega t)+\bm e_y E_\omega\sin\delta\sin\omega t,
\ee
where $\delta$ is the ellipticity. The value of $\delta=0$ corresponds to the linear polarization of the ac field along the $x$-axis, $\delta=\pm\pi/4$ -- to the left and right circular polarization, and $\delta=\pi/2$ -- to the linear polarization along the $y$-axis. Then, following the same steps as before we get the time averaged electric current $\bm j_0 =\bm e_x j_x^0$, where
\be 
\frac{ j_x^0}{en_sv_F} =
\int^{\infty}_0 e^{-\xi}  d \xi 
\frac 1{\pi}\int_{0}^\pi dx 
\left({\cal F}_0\xi+Z\cos\delta \cos x\right)
N\left(\sqrt{\left({\cal F}_0\xi+Z\cos\delta\cos x\right)^2
+ \left(Z\sin\delta\sin x\right)^2}\right).
\ee
This results does not evidently depend on the sign of $\delta$, i.e., on the direction (left or right) of the elliptic polarization. 
For any value of the ellipticity $\delta$ the current flows only in the direction of the dc electric field: the current component $j_y^0$ and the photoconductivity $\sigma_{yx}^{\rm ph}$ equal zero in the considered case. In the limit $E_0\to 0$ we then get
\be 
\frac{\sigma^{\rm ph}_{xx}({\cal F}_\omega,\omega\tau,\delta)}{\sigma_0} =
{\cal C}({\cal F}_\omega,\omega\tau,\delta)-\cos^2\delta
{\cal D}({\cal F}_\omega,\omega\tau,\delta)
\label{Pcond-DeltaXX}
\ee
where we have introduced two new functions
\be 
{\cal C}({\cal F}_\omega,\omega\tau,\delta)=\int^{\infty}_0 \xi e^{-\xi}  d \xi 
\frac 2{\pi}\int_{0}^{\pi/2} dx 
N\left(\sqrt{\left(Z\cos\delta\sin x\right)^2
+ \left(Z\sin\delta\cos x\right)^2}\right),
\label{C}
\ee
\be 
{\cal D}({\cal F}_\omega,\omega\tau,\delta)=\int^{\infty}_0 \xi e^{-\xi}  d \xi 
\frac 2{\pi}\int_{0}^{\pi/2} dx (Z\sin x)^2 
M\left(\sqrt{\left(Z\cos\delta\sin x\right)^2
+ \left(Z\sin\delta\cos x\right)^2}\right).
\label{D}
\ee
Comparing the definitions (\ref{C})--(\ref{D}) and (\ref{A})--(\ref{B}) we see that the following identities are valid:
\be 
{\cal C}({\cal F}_\omega,\omega\tau,0)={\cal C}({\cal F}_\omega,\omega\tau,\pi/2)={\cal A}({\cal F}_\omega,\omega\tau),\ \ 
{\cal D}({\cal F}_\omega,\omega\tau,0)={\cal B}({\cal F}_\omega,\omega\tau).
\ee 
Consequently, equation (\ref{Pcond-DeltaXX}) gives the same result at $\delta=0$ and $\delta=\pi/2$ as equation (\ref{Pcond-ThetaXX}) at $\theta=0$ and $\theta=\pi/2$. In the circular polarization case $\delta=\pi/4$ the argument of the functions $N$ and $M$ in Eqs. (\ref{C})--(\ref{D}) does not depend on $x$, the integral over $dx$ can be taken, and the formulas (\ref{C})--(\ref{D}) are simplified:
\be 
{\cal C}({\cal F}_\omega,\omega\tau,\pi/4)=
\int^{\infty}_0 \xi e^{-\xi}  d \xi 
N\left(\frac{Z}{\sqrt{2}}\right)=\int^{\infty}_0 \xi e^{-\xi} N\left(\frac{{\cal F}_\omega}{\sqrt{2}}  \frac{\sin(\omega\tau\xi/2)}{\omega \tau/2}\right) d \xi ,
\label{Cd}
\ee
\be 
{\cal D}({\cal F}_\omega,\omega\tau,\pi/4)=
\frac 12\int^{\infty}_0 \xi e^{-\xi} Z^2 M\left(\frac{Z}{\sqrt{2}}\right)d \xi 
=\frac 12{\cal F}_\omega^2 \int^{\infty}_0 \xi e^{-\xi} \left(\frac{\sin(\omega\tau\xi/2)}{\omega \tau/2}\right)^2M\left(\frac{{\cal F}_\omega}{\sqrt{2}}  \frac{\sin(\omega\tau\xi/2)}{\omega \tau/2}\right) d \xi .
\label{Dd}
\ee
 
\begin{figure}[ht]
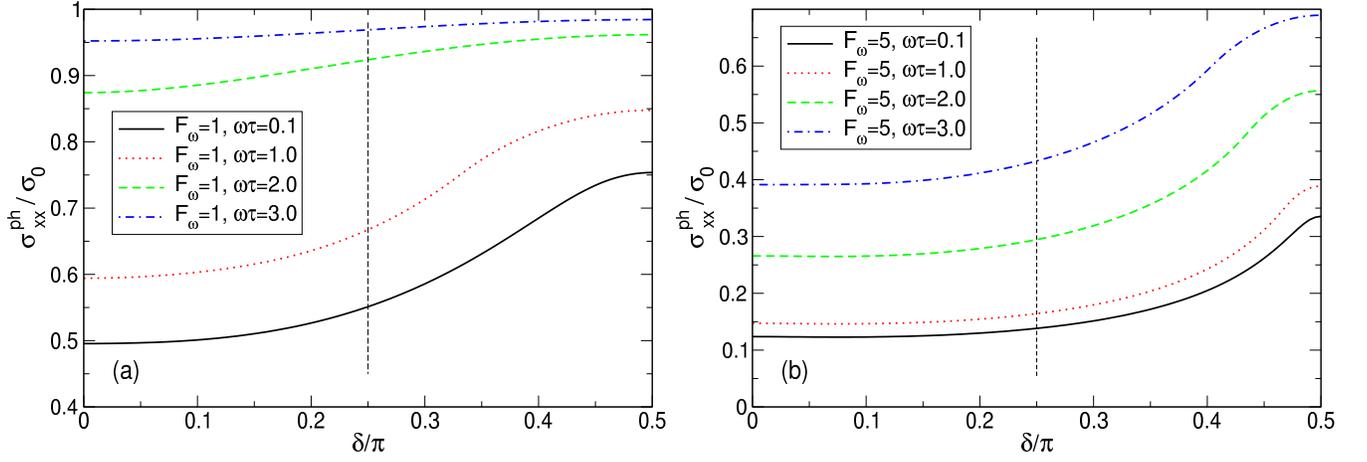

\includegraphics[width=0.49\textwidth]{SigmaXX-Delta_Fw1.eps}  
\includegraphics[width=0.49\textwidth]{SigmaXX-Delta_Fw5.eps}  
\caption{\label{fig:PCvsDeltaXX} The photoconductivity (\ref{Pcond-DeltaXX}) as a function of the ellipticity $\delta$ at several values of $\omega\tau$ and at (a) the electric field parameter ${\cal F}_\omega=1$ and (b) ${\cal F}_\omega=5$. The left and right edges of the plot, $\delta=0$ and $\delta=\pi/2$, correspond to linear polarizations of radiation along the $x$- and $y$-axis respectively. The central dashed line at $\delta=\pi/4$ refers to the circular polarization of radiation. }
\end{figure}

Figure \ref{fig:PCvsDeltaXX} shows the photoconductivity $\sigma^{\rm ph}_{xx}({\cal F}_\omega,\omega\tau,\delta)$ as a function of the ellipticity $\delta$ in the moderate (${\cal F}_\omega=1$) and strong (${\cal F}_\omega=5$) electric fields at a few values of $\omega\tau$. Qualitatively, the dependencies shown in Figure \ref{fig:PCvsDeltaXX} are similar to those on Figure \ref{fig:PCvsThetaXX}: the photoconductivity is quite strong already in moderate electric fields, is very sensitive to the ellipticity, and the influence of radiation of the conductivity is more essential at large electric fields and low frequencies. 

\subsection{Electric field and frequency dependence of the photoconductivity\label{sec:E&w-dependence}}

Now we analyze the photoconductivity dependencies on the electric field and frequency parameters ${\cal F}_\omega$ and $\omega\tau$. We consider several typical cases.

\subsubsection{Linear polarization, parallel orientation of the dc and ac fields; diagonal photoconductivity}

Here we consider the case of the linearly polarized radiation with the parallel polarizations of the dc and ac electric fields, $\delta=0$, $\theta=0$, $\bm E_0\parallel\bm E_\omega$. As we have seen in Section \ref{sec:LinearPolariz}, the photoconductivity effect is the largest in this case. Equation (\ref{Pcond-ThetaXX}) gives in this limit 
\be 
\frac{\sigma_{xx}^{\rm ph}({\cal F}_\omega,\omega\tau,0)}{\sigma_0} =
{\cal A}({\cal F}_\omega,\omega\tau)- 
{\cal B}({\cal F}_\omega,\omega\tau) \equiv
\frac{\sigma_0-\delta \sigma_{xx}^{\parallel}}{\sigma_0}.
\label{Pcond-ThetaXX-parallel}
\ee
Here we introduce the difference $\delta \sigma_{xx}^{\parallel}=\sigma_0-\sigma_{xx}^{\rm ph}({\cal F}_\omega,\omega\tau,0)$, to emphasize how the conductivity changes under the influence of radiation. Since the photoconductivity of graphene is negative, the function $\delta \sigma_{xx}^{\parallel}$ is larger than zero; the superscript $\parallel$ reminds that we are dealing with the parallel orientation of the dc and ac fields.   

\begin{figure}[ht]
\includegraphics[width=0.49\textwidth]{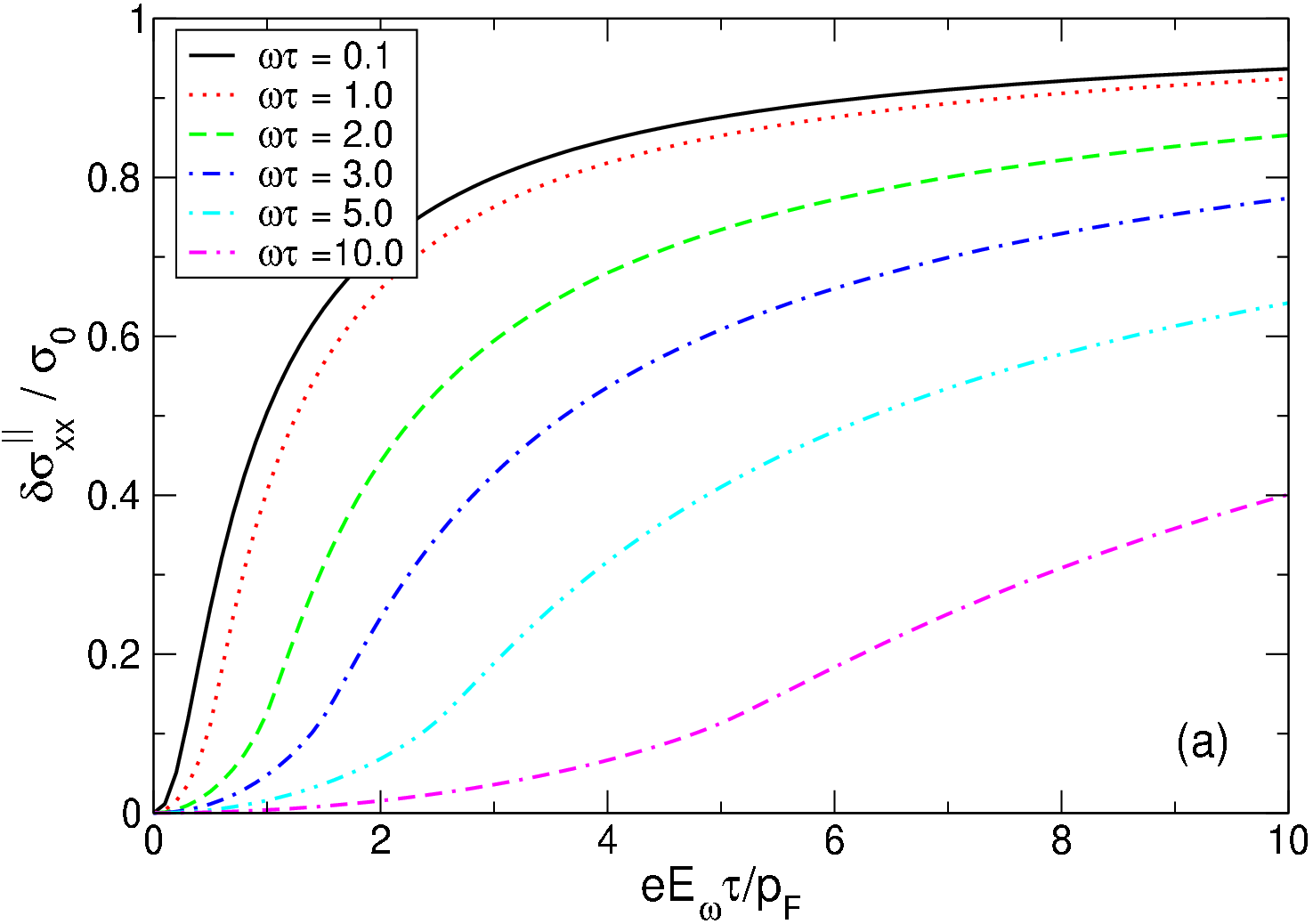}  
\includegraphics[width=0.49\textwidth]{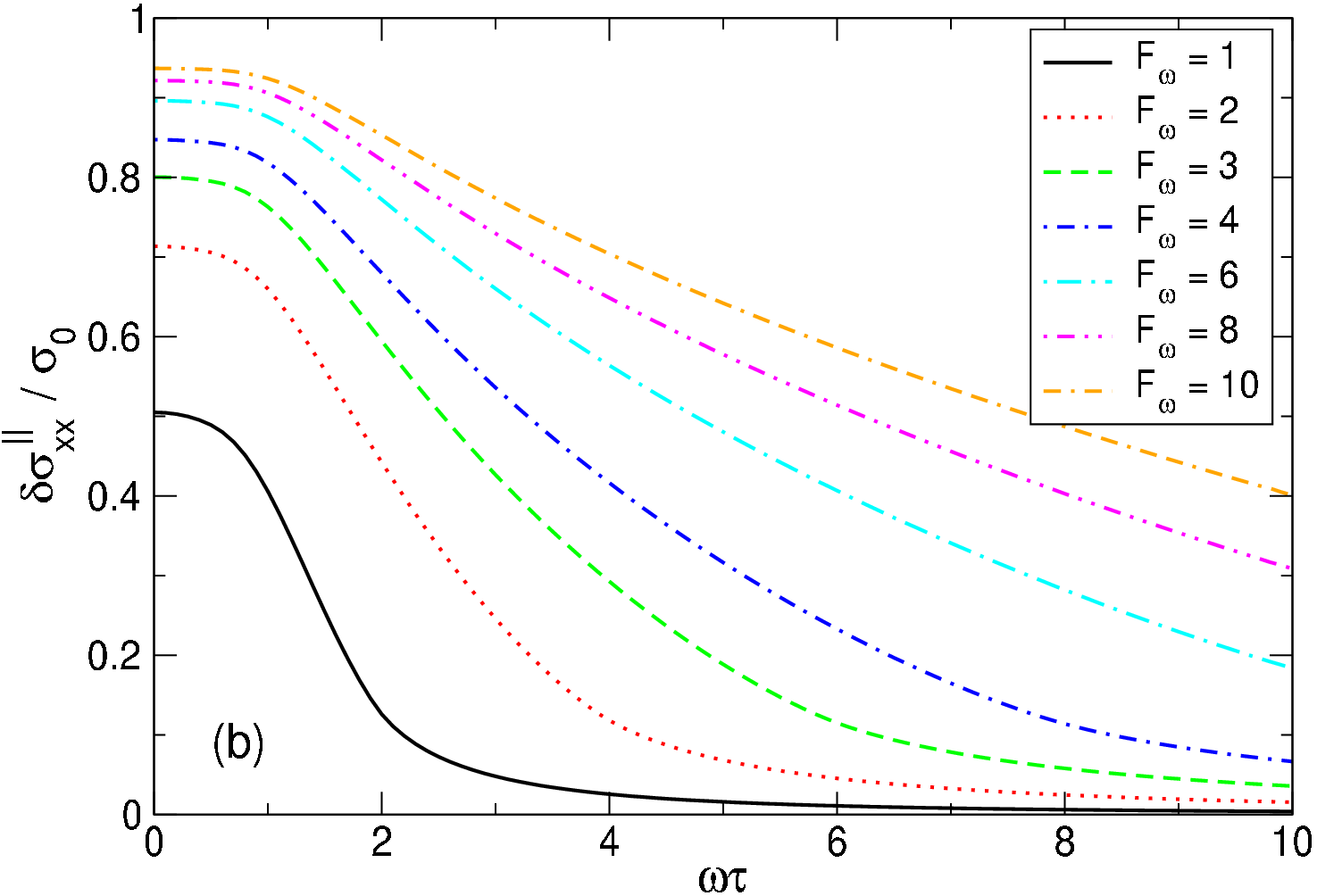}  
\caption{\label{fig:Spar} The photoconductivity change $\delta \sigma_{xx}^{\parallel}$, defined by Eq. (\ref{Pcond-ThetaXX-parallel}), as a function of (a) the electric field parameter ${\cal F}_\omega$ at fixed values of $\omega\tau$, and (b) the frequency parameter $\omega\tau$ at fixed values of ${\cal F}_\omega$.}
\end{figure}

Figure \ref{fig:Spar}(a) shows the field dependence of the function $\delta \sigma_{xx}^{\parallel}$ defined by Eq. (\ref{Pcond-ThetaXX-parallel}). When the field parameter ${\cal F}_\omega$ grows the function $\delta \sigma_{xx}^{\parallel}$ first quickly increases and then saturates. The saturation level of $\delta \sigma_{xx}^{\parallel}$ can be larger than $\sim 0.95\sigma_0$ at low frequencies $\omega\tau\lesssim 0.1$ and large ac electric fields ${\cal F}_\omega\gtrsim 10$. The boundary between the strong-growth and saturation intervals on the field axis lies at ${\cal F}_\omega\simeq 1$ at $\omega\tau\lesssim 1$ and at ${\cal F}_\omega\simeq\omega\tau$ at $\omega\tau\gtrsim 1$. Figure \ref{fig:Spar}(b) illustrates the frequency dependence of $\delta \sigma_{xx}^{\parallel}$. It falls down quite quickly with $\omega\tau$ and decreases by a factor of order two at $\omega\tau\simeq {\cal F}_\omega$.

\subsubsection{Linear polarization, orthogonal orientation of the dc and ac fields; diagonal photoconductivity}

\begin{figure}[ht]
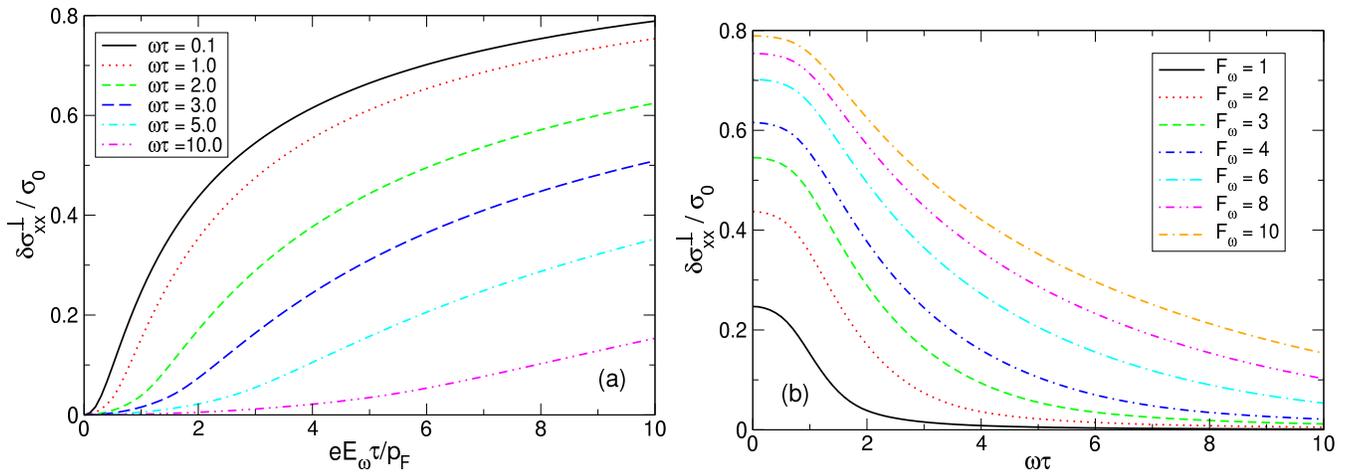

\includegraphics[width=0.49\textwidth]{SigmaXXper-Fw.eps}  
\includegraphics[width=0.49\textwidth]{SigmaXXper-wt.eps}  
\caption{\label{fig:Sper} The photoconductivity change $\delta \sigma_{xx}^{\perp}$ defined by Eq. (\ref{Pcond-ThetaXX-perpendicular}), as a function of (a) the electric field parameter ${\cal F}_\omega$ at fixed values of $\omega\tau$, and (b) the frequency parameter $\omega\tau$ at fixed values of ${\cal F}_\omega$.}
\end{figure}

Now we consider the case of perpendicular polarizations, $\delta=\pi/2$, $\theta=\pi/2$, $\bm E_0\parallel\bm e_x\perp \bm E_\omega\parallel\bm e_y$. Then we get from equation (\ref{Pcond-ThetaXX}) 
\be 
\frac{\sigma_{xx}^{\rm ph}({\cal F}_\omega,\omega\tau,\pi/2)}{\sigma_0} =
{\cal A}({\cal F}_\omega,\omega\tau)\equiv
\frac{\sigma_0-\delta \sigma_{xx}^{\perp}}{\sigma_0},
\label{Pcond-ThetaXX-perpendicular}
\ee
Figures \ref{fig:Sper}(a,b) show the field and frequency dependencies of the function $\delta \sigma_{xx}^{\perp}$ defined by Eq. (\ref{Pcond-ThetaXX-perpendicular}). The general trends of these dependencies is similar to those of the function $\delta \sigma_{xx}^{\parallel}$, but quantitatively, the conductivity change is weaker. The growth of $\delta \sigma_{xx}^{\perp}$ with the field is slower, Figure \ref{fig:Sper}(a), and its decrease with $\omega\tau$ is faster, Figure \ref{fig:Sper}(b), than for the parallel-polarization function $\delta \sigma_{xx}^{\parallel}$, Figure \ref{fig:Spar}. 

\subsubsection{Linear polarization, off-diagonal photoconductivity}

\begin{figure}[ht]
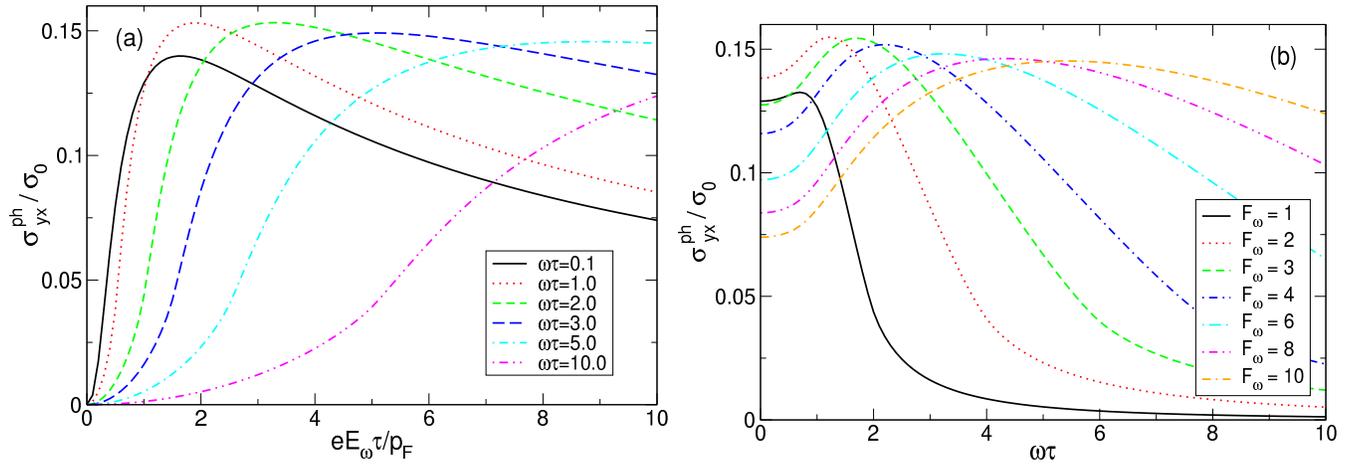

\includegraphics[width=0.49\textwidth]{SigmaYX-Fw.eps}  
\includegraphics[width=0.49\textwidth]{SigmaYX-wt.eps}  
\caption{\label{fig:Syx} The photoconductivity $\sigma_{yx}^{\rm ph}$ defined by Eq. (\ref{PcondYX-max}), at $\theta=3\pi/4$, as a function of (a) the electric field parameter ${\cal F}_\omega$ at fixed values of $\omega\tau$, and (b) the frequency parameter $\omega\tau$ at fixed values of ${\cal F}_\omega$.}
\end{figure}

The off-diagonal conductivity $\sigma_{yx}^{\rm ph}$ is determined by Eq. (\ref{Pcond-ThetaYX}). The maximum values of $|\sigma_{yx}^{\rm ph}|$ are reached at $\theta=\pi/2\pm \pi/4$, see Figure \ref{fig:PCvsThetaYX}, and are equal to 
\be 
\frac{\sigma_{yx}^{\rm ph}({\cal F}_\omega,\omega\tau,\pi/2\pm \pi/4)}{\sigma_0} =
\pm\frac 12 
{\cal B}({\cal F}_\omega,\omega\tau).
\label{PcondYX-max}
\ee
Figures \ref{fig:Syx}(a,b) show the electric field and frequency dependencies of the maximal off-diagonal photoconductivity (\ref{PcondYX-max}) at $\theta=3\pi/4$. These dependencies substantially differ from those of the diagonal photoconductivity. Both the field and frequency dependencies are non-monotonic and have a maximum. For any value of the frequency parameter, the photoconductivity $\sigma_{yx}^{\rm ph}$ first grow with the electric field, Figure \ref{fig:Syx}(a), reaches a maximum and then decreases. The maximum is at ${\cal F}_\omega\simeq 1$ for small frequencies $\omega\tau\lesssim 1$ and then approximately follows $\sim 2\omega\tau$ when the frequency increases. The frequency dependence also demonstrates a maximum of $\sigma_{yx}^{\rm ph}$ at $\omega\tau\simeq {\cal F}_\omega/2$, Figure \ref{fig:Syx}(b). The absolute value of $\sigma_{yx}^{\rm ph}$ in the maximum is about $0.16$, in units of $\sigma_0$. Thus, the irradiation of graphene by a linearly polarized electromagnetic wave at the angle $\pi/4$ to the direction of the dc current may substantially influence the direction of the current flow. For example, if $\omega\tau=0.1$ and $\theta=3\pi/4$, the current deviates from the $x$-direction by $12$ and $25$ degrees at ${\cal F}_\omega=1$ and 5 respectively, see Figures \ref{fig:PCvsThetaXX} and \ref{fig:PCvsThetaYX}. At larger frequencies the deviations are smaller: at $\omega\tau=3$ the corresponding numbers are 9 and 15 degrees (at ${\cal F}_\omega=1$ and 5). 

Physically, non-monotonic behavior of $\sigma_{yx}^{\rm ph}$ is explained by the competition of two factors. At low ac electric fields, the electron distribution function is isotropic, see Figure \ref{fig:distrfunc}(a) in Appendix \ref{app:distrfun}, and close to the equilibrium one: the current $j_y$ and the transverse photoconductivity $\sigma_{yx}^{\rm ph}$ are small. When the field parameter ${\cal F}_\omega$ increases, the anisotropy degree increases too, Figure \ref{fig:distrfunc}(b,c,d), and $\sigma_{yx}^{\rm ph}$ grows. However, at large values ${\cal F}_\omega\gg 1$ the occupation by electrons of the high-energy states leads to the growth of their ``effective mass'' $E/v_F^2$, and $\sigma_{yx}^{\rm ph}$ slowly falls down similar to $\sigma_{xx}^{\rm ph}$, as was discussed in Section \ref{sec:LinearPolariz}.

\subsubsection{Circular polarization of the ac field; diagonal photoconductivity}

\begin{figure}[ht]
\includegraphics[width=0.49\textwidth]{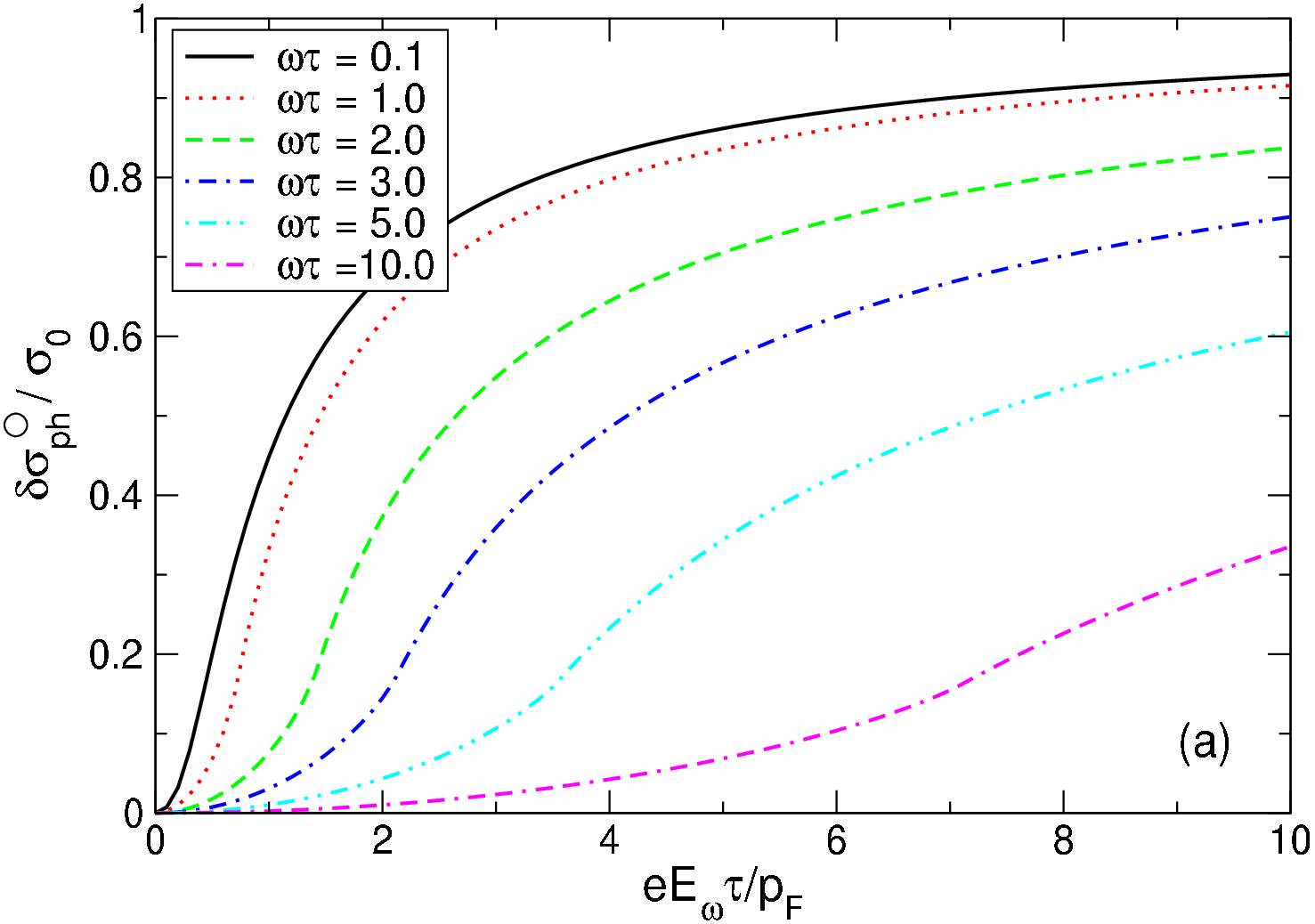}  
\includegraphics[width=0.49\textwidth]{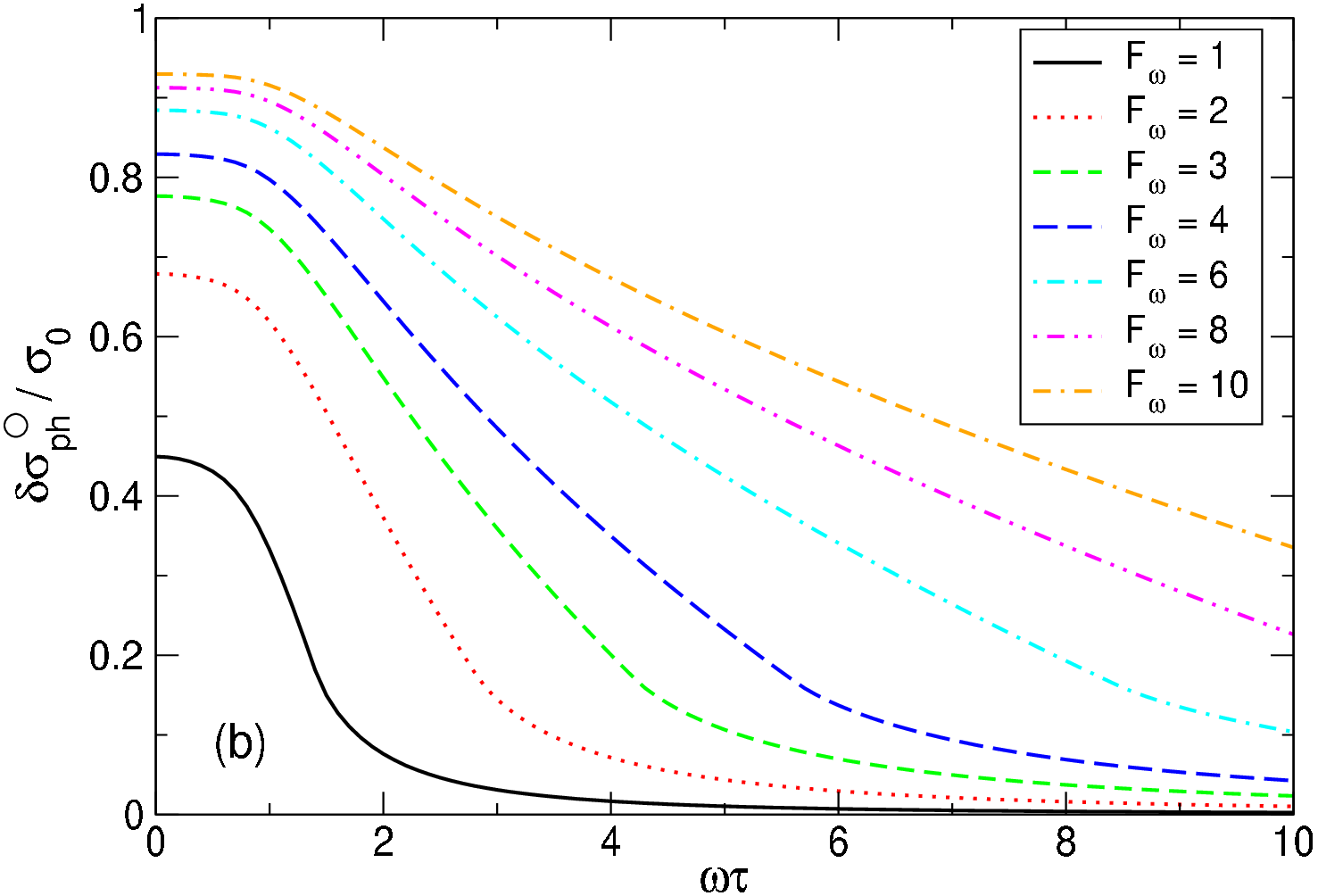}  
\caption{\label{fig:Scirc} The photoconductivity change $\delta\sigma^{\ocircle}_{xx}$ defined by Eq. (\ref{Pcond-DeltaXX-circular}) as a function of (a) the electric field parameter ${\cal F}_\omega$ at fixed values of $\omega\tau$, and (b) the frequency parameter $\omega\tau$ at fixed values of ${\cal F}_\omega$. }
\end{figure}

Finally we show results for the field and frequency dependencies of the diagonal photoconductivity $\delta\sigma_{xx}^{\ocircle}$ at the circular polarization of the incident radiation. In this case $\delta=\pi/4$ and we have from Eq. (\ref{Pcond-DeltaXX})
\be 
\frac{\sigma^{\rm ph}_{xx}({\cal F}_\omega,\omega\tau,\pi/4)}{\sigma_0} =
{\cal C}({\cal F}_\omega,\omega\tau,\pi/4)-\frac 12
{\cal D}({\cal F}_\omega,\omega\tau,\pi/4)\equiv \frac{\sigma_0-\delta\sigma^{\ocircle}_{xx} }{\sigma_0}.
\label{Pcond-DeltaXX-circular}
\ee
Figures \ref{fig:Scirc}(a) and (b) show the field and frequency dependencies of the photoconductivity $\delta\sigma^{\ocircle}_{xx}$ defined by Eq. (\ref{Pcond-DeltaXX-circular}). Qualitatively, the dependencies shown here are similar to those obtained for $\delta\sigma^{\parallel}_{xx}$, Figure \ref{fig:Spar}, and $\delta\sigma^{\perp}_{xx}$, Figure \ref{fig:Sper}, but there is a quantitative difference. Altogether, Figures \ref{fig:Spar} -- \ref{fig:Scirc} provide a complete picture of the field and frequency dependencies of the photoconductivity at different polarizations of the incident electromagnetic waves.  

\subsection{Comparison with the perturbation theory\label{sec:perturb-expansion}}

The perturbation theory results give a correction to the material conductivity proportional to the squared electric field. As seen from Figures \ref{fig:Spar}(a) -- \ref{fig:Scirc}(a), the exact result deviates from the ${\cal F}_\omega^2$ dependence at rather low values of the electric field parameter ${\cal F}_\omega$. It makes sense to compare quantitatively results of the third-order perturbation theory \cite{Mikhailov16a} with the exact results obtained here.

In the limit of low electric fields we expand the functions ${\cal A}$ -- ${\cal D}$ in equations (\ref{A}) -- (\ref{B}) and (\ref{C}) -- (\ref{D}) in powers of ${\cal F}_\omega$ up to the second order $\sim {\cal F}_\omega^2$. Taking into account that the first terms of Taylor's expansion of the functions $N(x)$ and $M(x)$ are $N\left( x\right)=1-x^2/8+\dots$ and $M(x)= 1/4+\dots$ and taking the integrals over $dx$ and $d\xi$ analytically, we find that the functions ${\cal C}$ and ${\cal D}$ do not depend on $\delta$ in the considered limit, all four functions are related to each other, ${\cal A}={\cal C}$, ${\cal B}={\cal D}$, ${\cal A}=1-{\cal B}/2$, and 
\be 
{\cal A}({\cal F}_\omega,\omega\tau)\approx
1 -\frac{{\cal F}_\omega^2}{8}\frac{3+(\omega\tau)^2}{(1+(\omega\tau)^2)^2}+O\left({\cal F}_\omega^4\right).
\ee
This gives the following results for the components of the photoconductivity tensor in the second order in ${\cal F}_\omega$:
\be 
\frac{\sigma_{xx}^{\rm ph}({\cal F}_\omega,\omega\tau,\theta)}{\sigma_0} =
1 -\frac{{\cal F}_\omega^2}{8}\frac{3+(\omega\tau)^2}{(1+(\omega\tau)^2)^2}
(1+2 \cos^2\theta)+O\left({\cal F}_\omega^4\right),
\label{SigmaXXasym}
\ee
\be 
\frac{\sigma_{yx}^{\rm ph}({\cal F}_\omega,\omega\tau,\theta)}{\sigma_0} =
- \frac {{\cal F}_\omega^2}8\frac{3+(\omega\tau)^2}{(1+(\omega\tau)^2)^2}
(2\sin\theta\cos\theta )+O\left({\cal F}_\omega^4\right);
\label{SigmaYXasym}
\ee
the asymptote of $\sigma^{\rm ph}_{xx}({\cal F}_\omega,\omega\tau,\delta)$ has the same form as (\ref{SigmaXXasym}), but with $\theta$ replaced by $\delta$. The results (\ref{SigmaXXasym}) -- (\ref{SigmaYXasym}) can be also derived from the general formulas of the third-order perturbation theory \cite{Mikhailov16a}, see Appendix \ref{app:perturbtheory}.

In Figure \ref{fig:Asy} we compare exact, nonperturbative theory curves with the low-field asymptotes (\ref{SigmaXXasym}) and (\ref{SigmaYXasym}). At low fields, roughly corresponding to the interval $0<{\cal F}_\omega\lesssim \omega\tau/2$,  the exact and approximate curves are close to each other. Then, in the interval  $\omega\tau/2\lesssim{\cal F}_\omega\lesssim \omega\tau$ the photoconductivities $\delta\sigma^{\parallel}_{xx}$ and $\sigma^{\rm ph}_{yx}$ calculated from exact formulas grow faster than the ${\cal F}_\omega^2$-approximations. Finally, at ${\cal F}_\omega\gtrsim \omega\tau$, the exact formulas exhibit saturation of $\delta\sigma^{\parallel}_{xx}$ and a tendency to the reduction of $\sigma^{\rm ph}_{yx}$, and the asymptotic formulas (\ref{SigmaXXasym}) and (\ref{SigmaYXasym}) become fully unreliable. In the limit of low frequencies $\omega\tau\lesssim 1$ the frequency $\omega$ should be replaced by $1/\tau$ in these estimates. The applicability of the third-order perturbation theory is thus restricted by the condition 
\be 
\frac{eE_\omega\tau}{\max\{1,\omega\tau\} p_F}\lesssim \frac 12.
\label{applicability}
\ee 
If, for example, the relaxation time $\tau\simeq 1$ ps, the frequency $f\simeq 1$ THz, and the electron density is $n_s\simeq 10^{11}$ cm$^{-2}$, the conditions $\omega\tau\gg 1$ and $\hbar\omega\lesssim 2E_F$ are satisfied, and the formula (\ref{applicability}) restricts the value of the ac electric field by $E_\omega\simeq 2.2$ kV/cm. At higher fields the nonperturbative theory should be applied. 

\begin{figure}[ht]
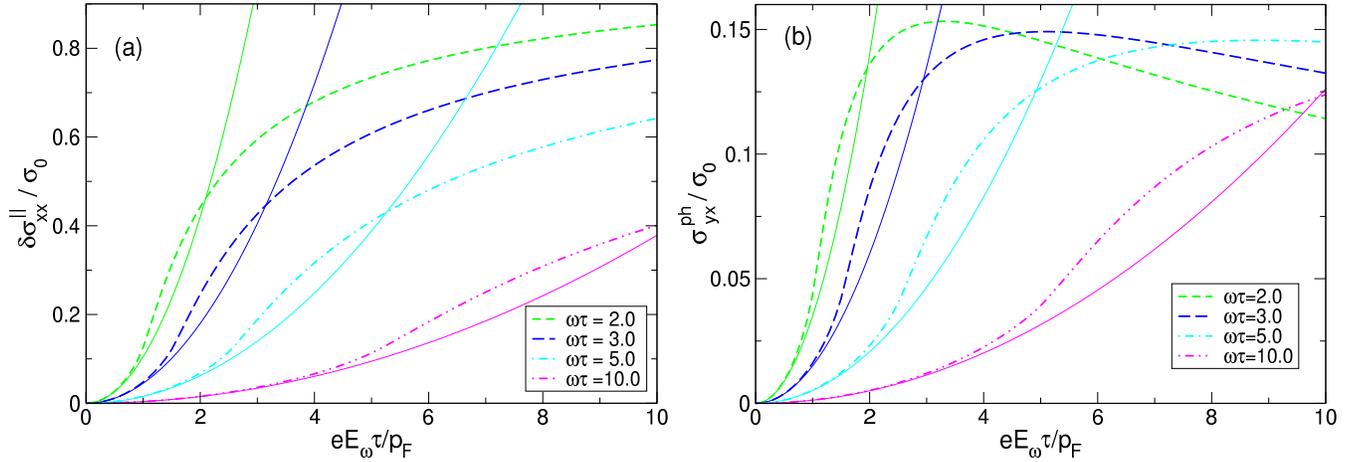

\includegraphics[width=0.49\textwidth]{SigmaXXpar-Fw_asy.eps}  
\includegraphics[width=0.49\textwidth]{SigmaYX-Fw_asy.eps}  
\caption{\label{fig:Asy} Comparison of exact (thick dashed or dot-dashed curves) and asymptotic (thin solid curves) formulas for the electric field dependencies of the photoconductivities (a) $\delta\sigma^{\parallel}_{xx}$ and (b) $\sigma^{\rm ph}_{yx}$ at several values of $\omega\tau$. }
\end{figure}

\subsection{Influence of temperature\label{sec:temperature}}

\begin{figure}[ht]
\includegraphics[width=0.49\textwidth]{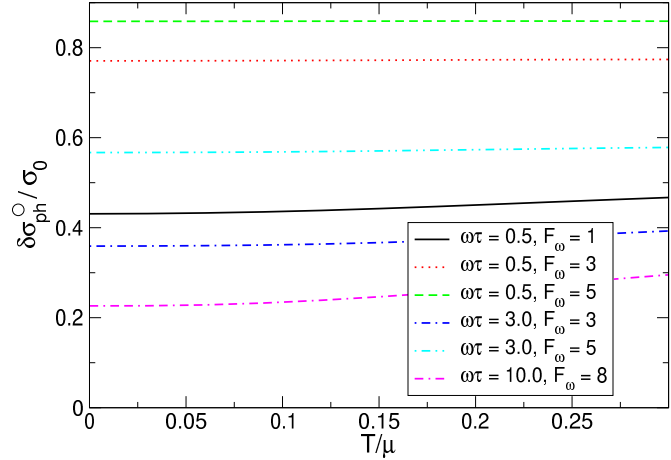}  
\caption{\label{fig:Scirc-Tmu} Temperature dependence of $\delta\sigma^{\ocircle}_{xx}$ at several values of ${\cal F}_\omega$ and $\omega\tau$. $\mu$ is the chemical potential.}
\end{figure}

So far we have used the simplified expression for the radiation induced current (\ref{genresforjT0}) which is valid at $T=0$. At a finite temperature $T>0$ any photoconductivity discussed above can be calculated using the relation
\ba 
\sigma\left(\mu,T\right)=
\frac 1{4T}\int_{-\infty}^\infty \frac{dE_F}{\cosh^2\left(\frac{\mu-E_F}{2T}\right)} \sigma\left(E_F,0\right)
\ea
between the finite-$T$ and zero-$T$ response functions; here $\mu$ is the chemical potential at $T\neq 0$. As we have mentioned in Section \ref{sec:GenFormulas} our results should not be very sensitive to $T$ since within the quasiclassical theory the case of intrinsic graphene ($E_F\approx 0$) is excluded. Figure \ref{fig:Scirc-Tmu} confirms this statement. It shows, as a representative example, the temperature dependence of the conductivity change $\delta\sigma^{\ocircle}_{xx}$, induced by a circularly polarized radiation, as a function of $T/\mu$ at several values of the electric field strength and frequency. One sees that $\delta\sigma^{\ocircle}_{xx}$ varies with the temperature by only a few percent when the parameter $T/\mu$ grows from zero up to $T/\mu=0.3$. At the electron density $\simeq 10^{12}-10^{13}$ cm$^{-2}$ the value of $T\approx 0.3 \mu$ corresponds to $\simeq 405-1280$ K. Therefore all our results shown in previous Sections give reliable estimates of the discussed physical effects both at room temperature $T_0$ and in the case when the electron gas in graphene is heated by the radiation up to $T\simeq (2-4)T_0$.

\section{Summary and conclusions\label{sec:summary}}

To summarize, we have developed a nonperturbative theory of graphene photoconductivity applicable at low ($\hbar\omega\lesssim 2E_F$) frequencies which, dependent on the electron density in the material, may cover the range from microwave up to near-infrared frequencies. We have investigated the dependencies of the photoconductivity tensor on all relevant physical parameters (electric field strength, frequency, temperature, material properties, etc.), and found the applicability boundaries of the third-order perturbation theory. We have shown that the photoconductivity effect strongly depends on the radiation frequency, being the largest at $\omega\tau\lesssim 1$, and that the conductivity change caused by the irradiation can be as large as $80-90$\% in quite moderate electric fields of order of kV/cm. We have also shown that the effect is very sensitive to the direction and/or the ellipticity of the electric field polarization of the incident electromagnetic radiation. The predicted dependencies can be used for detection of THz, far- and mid-infrared radiation. Our findings may be interesting for further fundamental experimental studies of the nonlinear electrodynamic effects in graphene, as well as for its applications in the field of nonlinear terahertz and infrared photonics and optoelectronics.

\begin{acknowledgments}
This work has received funding from the European Union's Horizon 2020 research and innovation programme Graphene Core 3 under Grant Agreement No. 881603.
\end{acknowledgments}

\appendix
\section{The distribution function of electrons under strong electromagnetic irradiation\label{app:distrfun}}

\begin{figure}[ht]
\includegraphics[width=0.49\textwidth]{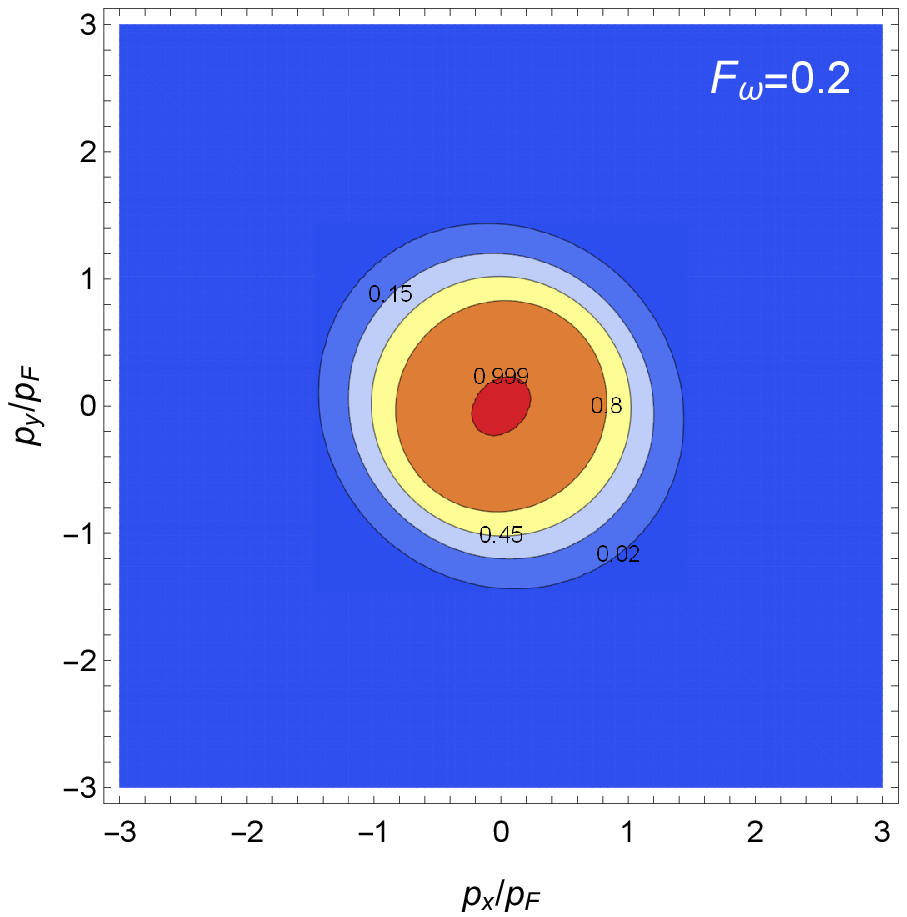}  
\includegraphics[width=0.49\textwidth]{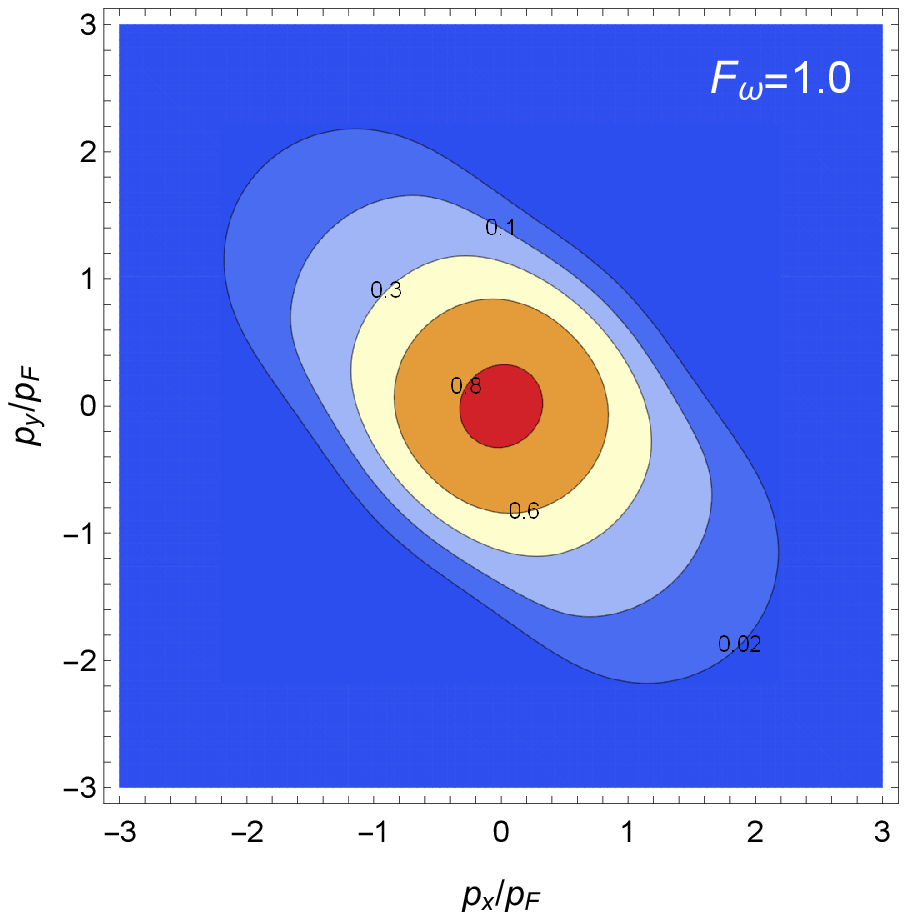}  
\includegraphics[width=0.49\textwidth]{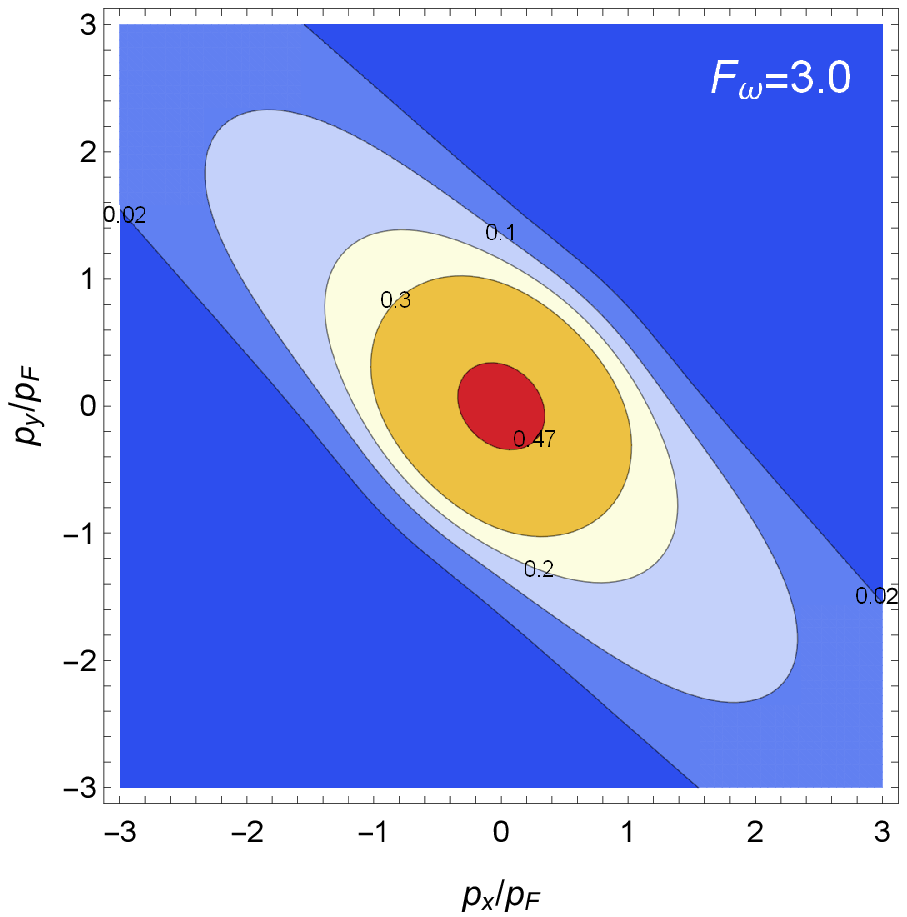}  
\includegraphics[width=0.49\textwidth]{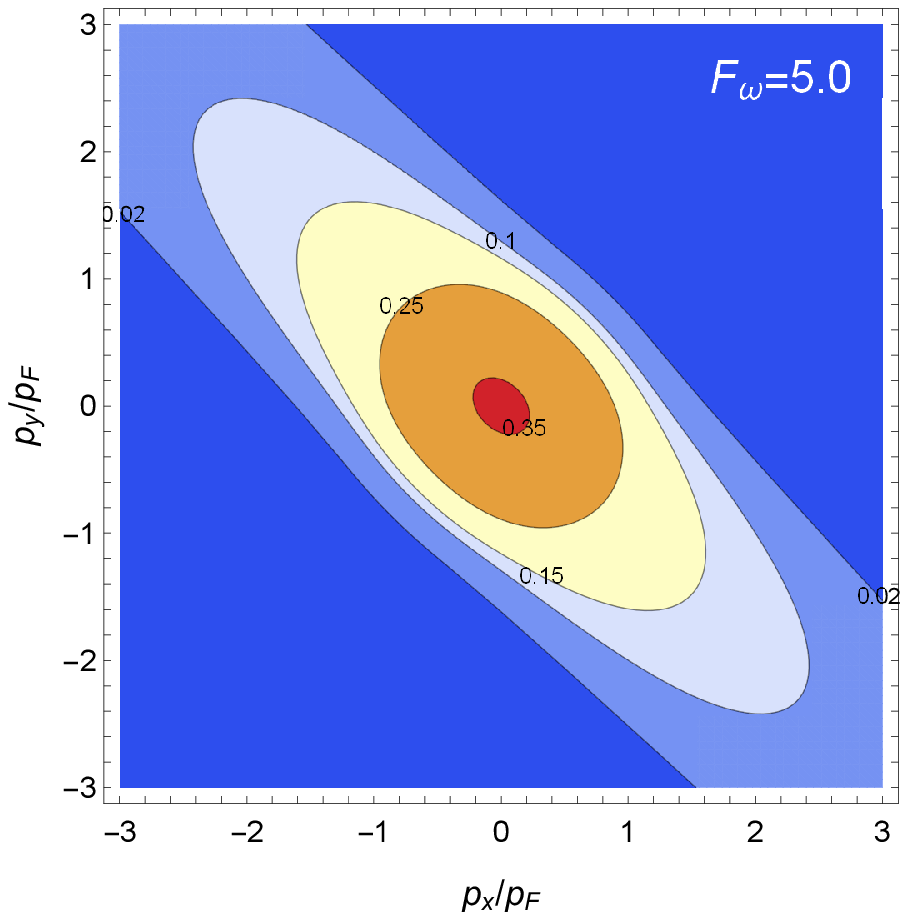}  
\caption{\label{fig:distrfunc} The time-averaged electron distribution function (\ref{solution}) under electromagnetic irradiation at ${\cal F}_\omega=0.2$, 1, 3, and 5. Other parameters are: $\omega\tau=1$, $T/\mu=0.1$, the polarization angle $\theta=3\pi/4$.}
\end{figure}

Figure \ref{fig:distrfunc} illustrates the time-averaged electron distribution function (\ref{solution}) in the momentum space, under the action of the strong linearly polarized ac electric field (the polarization angle $\theta=3\pi/4$). 
The distribution function becomes strongly elongated along the direction of the field at ${\cal F}_\omega\gtrsim 1$, which leads to anisotropic transport properties of the system. In particular, the average electron effective mass in the longitudinal direction (along the field) becomes heavier than in the transverse direction.  

\section{The photoconductivity from the third-order perturbation theory\label{app:perturbtheory}}

Here we derive formulas for graphene photoconductivity from the general results of the third-order perturbation theory \cite{Mikhailov16a}. This allows to find the relation between the components of the photoconductivity tensor $\sigma_{xx}^{\rm ph}$, $\sigma_{yx}^{\rm ph}$, derived from the nonperturbative theory in this work, and the third-order fourth-rank conductivity tensor $\sigma_{\alpha\beta\gamma\delta}^{(3)}(\omega_1,\omega_2,\omega_3)$ determined in the perturbation theory \cite{Mikhailov16a,Cheng15}.

In the perturbation theory the third-order current is determined by Eq. (\ref{Taylor}). In the case of the photoconductivity effect the Fourier component of the electric field ${\bm E}_{\omega_1}$ is written as
\be 
{\bm E}_{\omega_1} =\bm E_0\delta(\omega_1)+\frac{E_\omega}{2}\left[\bm I\delta(\omega_1-\omega)+\bm I^\star\delta(\omega_1+\omega)\right]
\label{FourierFields}
\ee
where $E_\omega$ is the real amplitude of the ac field and the vector $\bm I$ is the unit vector (maybe complex) which describes its polarization. For example, for the cases of linearly and elliptically polarized fields considered in Sections \ref{sec:LinearPolariz} and \ref{sec:ElliptPolariz} above the vectors $\bm I$ have the form $\bm I=(\cos\theta,\sin\theta)$ and $\bm I=(\cos\delta,i\sin\delta)$, respectively. Substituting the Fourier components (\ref{FourierFields}) in (\ref{Taylor}) we obtain, after some transformations, the time averaged third-order contribution to the current
\be
j_{\alpha}^{(3)}=
6\sigma_{\alpha\beta\gamma\delta}^{(3)}(0,\omega,-\omega) E_0^\beta \left(\frac{E_\omega}{2}\right)^2
I^\gamma (I^\delta)^\star .\label{perturb-current}
\ee
According to Ref. \cite{Mikhailov16a}, the main contribution to the third-order conductivity at low frequencies, i.e. in the quasiclassical limit when the interband transitions can be neglected, is given by the contribution $\sigma_{\alpha\beta\gamma\delta}^{(3/0)}$, see Eq. (62) in \cite{Mikhailov16a}. Symmetrizing the expression for $\sigma_{\alpha\beta\gamma\delta}^{(3)} (\omega_1,\omega_2,\omega_3)$ according to Eqs. (59) and (65) in Ref. \cite{Mikhailov16a} we get the following formula for the third-order conductivity in the quasiclassical limit
\ba 
\sigma_{\alpha\beta\gamma\delta}^{(3)} (\omega_1,\omega_2,\omega_3)&\approx &
\sigma_{\alpha\beta\gamma\delta}^{(3/0)} (\omega_1,\omega_2,\omega_3)
\nonumber \\ &=&
\frac 1{3!}\sigma_0^{(3)}
\frac {i\Delta_{\alpha\beta\gamma \delta} }{ \Omega_1+\Omega_2+\Omega_3+i\Gamma}
\Bigg[
\frac {1}{\Omega_1+\Omega_2+i\Gamma}\left( \frac {1}{\Omega_1+i\Gamma}+\frac {1}{\Omega_2+i\Gamma}\right)\nonumber \\ &+&
\frac {1}{\Omega_1+\Omega_3+i\Gamma}\left(\frac {1}{\Omega_1+i\Gamma}+\frac {1}{\Omega_3+i\Gamma}\right)+
\frac {1}{\Omega_2+\Omega_3+i\Gamma}\left(\frac {1}{\Omega_2+i\Gamma}+\frac {1}{\Omega_3+i\Gamma}\right)
\Bigg],\label{Sigma(3)asymp}
\ea
where $\Omega_i=\hbar\omega_i/E_F$, $\Gamma=\hbar/\tau E_F$, $\Delta_{\alpha\beta\gamma \delta}= \delta_{\alpha\beta}\delta_{\gamma \delta}+
\delta_{\alpha\gamma }\delta_{\beta\delta}+
\delta_{\alpha\delta}\delta_{\beta\gamma}$, and 
\be 
\sigma_0^{(3)}=\frac {e^4g_sg_v\hbar v_F^2}{16\pi E_F^4}
\label{sigma^30}.
\ee 
The asymptotic formula (82) in Ref. \cite{Mikhailov16a} follows from (\ref{Sigma(3)asymp}) under the conditions $\Gamma\ll|\Omega_i|\ll 1$ for all $i=1,2,3$. For the photoconductivity case we need the function $\sigma_{\alpha\beta\gamma\delta}^{(3)} (\omega_1,\omega_2,\omega_3)$ with the arguments $(\omega_1,\omega_2,\omega_3)=(0,\omega,-\omega)$. Then Eq. (\ref{Sigma(3)asymp}) gives
\ba 
\sigma_{\alpha\beta\gamma\delta}^{(3)} (0,\omega,-\omega)&=&
-\frac 1{3!}\sigma_0^{(3)}
\frac {\Delta_{\alpha\beta\gamma \delta} }{ \Gamma}
\frac {2(\Omega^2+3\Gamma^2)}{(\Omega^2+\Gamma^2)^2}=
-\sigma_0\frac {e^2\tau^2 }{12p_F^2}
\Delta_{\alpha\beta\gamma \delta} 
\frac {3+(\omega\tau)^2)}{(1+(\omega\tau)^2)^2}
\label{sigma(3)}
\ea
Substituting (\ref{sigma(3)}) in the photocurrent (\ref{perturb-current}) and adding the first-order contribution we get
\ba
j_{\alpha}=
\sigma_0\left\{\delta_{\alpha\beta}
-\frac {{\cal F}_\omega^2}{8}
\frac {(3+\omega^2\tau^2)}{(1+\omega^2\tau^2)^2}
\Big[\delta_{\alpha\beta}   +  I^\alpha (I^\beta)^\star +  (I^\alpha)^\star I^\beta \Big]\right\}E_0^\beta .\label{perturb-current2}
\ea
The tensor $P_{\alpha\beta}=\delta_{\alpha\beta}   +  I^\alpha (I^\beta)^\star +  (I^\alpha)^\star I^\beta $ determines the polarization dependence of the photoconductivity. For the linearly and elliptically polarized lights considered in Sections \ref{sec:LinearPolariz} and \ref{sec:ElliptPolariz} it equals
\be 
P_{\alpha\beta}^{\rm lin} =\left(\begin{array}{cc}
1+2\cos^2\theta & 2\sin\theta\cos\theta \\ 
2\sin\theta\cos\theta & 1+2\sin^2\theta \\
\end{array}
\right)\label{polariz-lin}
\ee
and 
\be 
P_{\alpha\beta}^{\rm ell} =\left(\begin{array}{cc}
1+2\cos^2\delta & 0 \\ 
0 & 1+2\sin^2\delta \\
\end{array}
\right)\label{polariz-ell}
\ee
respectively. Equation (\ref{perturb-current2}), together with (\ref{polariz-lin}) and (\ref{polariz-ell}), gives the result coinciding with Eqs. (\ref{SigmaXXasym}) -- (\ref{SigmaYXasym}) obtained by the Taylor expansion of the non-perturbative formulas in Section \ref{sec:perturb-expansion}. 

%


\begin{thebibliography}{30}%
\makeatletter
\providecommand \@ifxundefined [1]{%
 \@ifx{#1\undefined}
}%
\providecommand \@ifnum [1]{%
 \ifnum #1\expandafter \@firstoftwo
 \else \expandafter \@secondoftwo
 \fi
}%
\providecommand \@ifx [1]{%
 \ifx #1\expandafter \@firstoftwo
 \else \expandafter \@secondoftwo
 \fi
}%
\providecommand \natexlab [1]{#1}%
\providecommand \enquote  [1]{``#1''}%
\providecommand \bibnamefont  [1]{#1}%
\providecommand \bibfnamefont [1]{#1}%
\providecommand \citenamefont [1]{#1}%
\providecommand \href@noop [0]{\@secondoftwo}%
\providecommand \href [0]{\begingroup \@sanitize@url \@href}%
\providecommand \@href[1]{\@@startlink{#1}\@@href}%
\providecommand \@@href[1]{\endgroup#1\@@endlink}%
\providecommand \@sanitize@url [0]{\catcode `\\12\catcode `\$12\catcode
  `\&12\catcode `\#12\catcode `\^12\catcode `\_12\catcode `\%12\relax}%
\providecommand \@@startlink[1]{}%
\providecommand \@@endlink[0]{}%
\providecommand \url  [0]{\begingroup\@sanitize@url \@url }%
\providecommand \@url [1]{\endgroup\@href {#1}{\urlprefix }}%
\providecommand \urlprefix  [0]{URL }%
\providecommand \Eprint [0]{\href }%
\providecommand \doibase [0]{https://doi.org/}%
\providecommand \selectlanguage [0]{\@gobble}%
\providecommand \bibinfo  [0]{\@secondoftwo}%
\providecommand \bibfield  [0]{\@secondoftwo}%
\providecommand \translation [1]{[#1]}%
\providecommand \BibitemOpen [0]{}%
\providecommand \bibitemStop [0]{}%
\providecommand \bibitemNoStop [0]{.\EOS\space}%
\providecommand \EOS [0]{\spacefactor3000\relax}%
\providecommand \BibitemShut  [1]{\csname bibitem#1\endcsname}%
\let\auto@bib@innerbib\@empty
\bibitem [{\citenamefont {Castro~Neto}\ \emph {et~al.}(2009)\citenamefont
  {Castro~Neto}, \citenamefont {Guinea}, \citenamefont {Peres}, \citenamefont
  {Novoselov},\ and\ \citenamefont {Geim}}]{Neto09}%
  \BibitemOpen
  \bibfield  {author} {\bibinfo {author} {\bibfnamefont {A.~H.}\ \bibnamefont
  {Castro~Neto}}, \bibinfo {author} {\bibfnamefont {F.}~\bibnamefont {Guinea}},
  \bibinfo {author} {\bibfnamefont {N.~M.~R.}\ \bibnamefont {Peres}}, \bibinfo
  {author} {\bibfnamefont {K.~S.}\ \bibnamefont {Novoselov}},\ and\ \bibinfo
  {author} {\bibfnamefont {A.~K.}\ \bibnamefont {Geim}},\ }\bibfield  {title}
  {\bibinfo {title} {The electronic properties of graphene},\ }\href@noop {}
  {\bibfield  {journal} {\bibinfo  {journal} {Rev. Mod. Phys.}\ }\textbf
  {\bibinfo {volume} {81}},\ \bibinfo {pages} {109} (\bibinfo {year}
  {2009})}\BibitemShut {NoStop}%
\bibitem [{\citenamefont {Wallace}(1947)}]{Wallace47}%
  \BibitemOpen
  \bibfield  {author} {\bibinfo {author} {\bibfnamefont {P.~R.}\ \bibnamefont
  {Wallace}},\ }\bibfield  {title} {\bibinfo {title} {The band theory of
  graphite},\ }\href@noop {} {\bibfield  {journal} {\bibinfo  {journal} {Phys.
  Rev.}\ }\textbf {\bibinfo {volume} {71}},\ \bibinfo {pages} {622} (\bibinfo
  {year} {1947})}\BibitemShut {NoStop}%
\bibitem [{\citenamefont {Mikhailov}(2007)}]{Mikhailov07e}%
  \BibitemOpen
  \bibfield  {author} {\bibinfo {author} {\bibfnamefont {S.~A.}\ \bibnamefont
  {Mikhailov}},\ }\bibfield  {title} {\bibinfo {title} {Non-linear
  electromagnetic response of graphene},\ }\href@noop {} {\bibfield  {journal}
  {\bibinfo  {journal} {Europhys. Lett.}\ }\textbf {\bibinfo {volume} {79}},\
  \bibinfo {pages} {27002} (\bibinfo {year} {2007})}\BibitemShut {NoStop}%
\bibitem [{\citenamefont {Mikhailov}\ and\ \citenamefont
  {Ziegler}(2008)}]{Mikhailov08a}%
  \BibitemOpen
  \bibfield  {author} {\bibinfo {author} {\bibfnamefont {S.~A.}\ \bibnamefont
  {Mikhailov}}\ and\ \bibinfo {author} {\bibfnamefont {K.}~\bibnamefont
  {Ziegler}},\ }\bibfield  {title} {\bibinfo {title} {Non-linear
  electromagnetic response of graphene: {F}requency multiplication and the
  self-consistent field effects},\ }\href@noop {} {\bibfield  {journal}
  {\bibinfo  {journal} {J. Phys. Condens. Matter}\ }\textbf {\bibinfo {volume}
  {20}},\ \bibinfo {pages} {384204} (\bibinfo {year} {2008})}\BibitemShut
  {NoStop}%
\bibitem [{\citenamefont {Hendry}\ \emph {et~al.}(2010)\citenamefont {Hendry},
  \citenamefont {Hale}, \citenamefont {Moger}, \citenamefont {Savchenko},\ and\
  \citenamefont {Mikhailov}}]{Hendry10}%
  \BibitemOpen
  \bibfield  {author} {\bibinfo {author} {\bibfnamefont {E.}~\bibnamefont
  {Hendry}}, \bibinfo {author} {\bibfnamefont {P.~J.}\ \bibnamefont {Hale}},
  \bibinfo {author} {\bibfnamefont {J.~J.}\ \bibnamefont {Moger}}, \bibinfo
  {author} {\bibfnamefont {A.~K.}\ \bibnamefont {Savchenko}},\ and\ \bibinfo
  {author} {\bibfnamefont {S.~A.}\ \bibnamefont {Mikhailov}},\ }\bibfield
  {title} {\bibinfo {title} {Coherent nonlinear optical response of graphene},\
  }\href@noop {} {\bibfield  {journal} {\bibinfo  {journal} {Phys. Rev. Lett.}\
  }\textbf {\bibinfo {volume} {105}},\ \bibinfo {pages} {097401} (\bibinfo
  {year} {2010})}\BibitemShut {NoStop}%
\bibitem [{\citenamefont {Bykov}\ \emph {et~al.}(2012)\citenamefont {Bykov},
  \citenamefont {Murzina}, \citenamefont {Rybin},\ and\ \citenamefont
  {Obraztsova}}]{Bykov12}%
  \BibitemOpen
  \bibfield  {author} {\bibinfo {author} {\bibfnamefont {A.~Y.}\ \bibnamefont
  {Bykov}}, \bibinfo {author} {\bibfnamefont {T.~V.}\ \bibnamefont {Murzina}},
  \bibinfo {author} {\bibfnamefont {M.~G.}\ \bibnamefont {Rybin}},\ and\
  \bibinfo {author} {\bibfnamefont {E.~D.}\ \bibnamefont {Obraztsova}},\
  }\bibfield  {title} {\bibinfo {title} {Second harmonic generation in
  multilayer graphene induced by direct electric current},\ }\href@noop {}
  {\bibfield  {journal} {\bibinfo  {journal} {Phys. Rev. B}\ }\textbf {\bibinfo
  {volume} {85}},\ \bibinfo {pages} {121413(R)} (\bibinfo {year}
  {2012})}\BibitemShut {NoStop}%
\bibitem [{\citenamefont {Mikhailov}(2011)}]{Mikhailov11c}%
  \BibitemOpen
  \bibfield  {author} {\bibinfo {author} {\bibfnamefont {S.~A.}\ \bibnamefont
  {Mikhailov}},\ }\bibfield  {title} {\bibinfo {title} {Theory of the giant
  plasmon-enhanced second-harmonic generation in graphene and semiconductor
  two-dimensional electron systems},\ }\href@noop {} {\bibfield  {journal}
  {\bibinfo  {journal} {Phys. Rev. B}\ }\textbf {\bibinfo {volume} {84}},\
  \bibinfo {pages} {045432} (\bibinfo {year} {2011})}\BibitemShut {NoStop}%
\bibitem [{\citenamefont {Cheng}\ \emph
  {et~al.}(2014{\natexlab{a}})\citenamefont {Cheng}, \citenamefont
  {Vermeulen},\ and\ \citenamefont {Sipe}}]{Cheng14a}%
  \BibitemOpen
  \bibfield  {author} {\bibinfo {author} {\bibfnamefont {J.~L.}\ \bibnamefont
  {Cheng}}, \bibinfo {author} {\bibfnamefont {N.}~\bibnamefont {Vermeulen}},\
  and\ \bibinfo {author} {\bibfnamefont {J.~E.}\ \bibnamefont {Sipe}},\
  }\bibfield  {title} {\bibinfo {title} {Third order optical nonlinearity of
  graphene},\ }\href@noop {} {\bibfield  {journal} {\bibinfo  {journal} {New J.
  Phys.}\ }\textbf {\bibinfo {volume} {16}},\ \bibinfo {pages} {053014}
  (\bibinfo {year} {2014}{\natexlab{a}})}\BibitemShut {NoStop}%
\bibitem [{\citenamefont {Cheng}\ \emph
  {et~al.}(2014{\natexlab{b}})\citenamefont {Cheng}, \citenamefont
  {Vermeulen},\ and\ \citenamefont {Sipe}}]{Cheng14b}%
  \BibitemOpen
  \bibfield  {author} {\bibinfo {author} {\bibfnamefont {J.~L.}\ \bibnamefont
  {Cheng}}, \bibinfo {author} {\bibfnamefont {N.}~\bibnamefont {Vermeulen}},\
  and\ \bibinfo {author} {\bibfnamefont {J.~E.}\ \bibnamefont {Sipe}},\
  }\bibfield  {title} {\bibinfo {title} {Dc current induced second order
  optical nonlinearity in graphene},\ }\href@noop {} {\bibfield  {journal}
  {\bibinfo  {journal} {Optics Express}\ }\textbf {\bibinfo {volume} {22}},\
  \bibinfo {pages} {15868} (\bibinfo {year} {2014}{\natexlab{b}})}\BibitemShut
  {NoStop}%
\bibitem [{\citenamefont {Cheng}\ \emph {et~al.}(2015)\citenamefont {Cheng},
  \citenamefont {Vermeulen},\ and\ \citenamefont {Sipe}}]{Cheng15}%
  \BibitemOpen
  \bibfield  {author} {\bibinfo {author} {\bibfnamefont {J.~L.}\ \bibnamefont
  {Cheng}}, \bibinfo {author} {\bibfnamefont {N.}~\bibnamefont {Vermeulen}},\
  and\ \bibinfo {author} {\bibfnamefont {J.~E.}\ \bibnamefont {Sipe}},\
  }\bibfield  {title} {\bibinfo {title} {Third-order nonlinearity of graphene:
  Effects of phenomenological relaxation and finite temperature},\ }\href@noop
  {} {\bibfield  {journal} {\bibinfo  {journal} {Phys. Rev. B}\ }\textbf
  {\bibinfo {volume} {91}},\ \bibinfo {pages} {235320} (\bibinfo {year}
  {2015})}\BibitemShut {NoStop}%
\bibitem [{\citenamefont {Mikhailov}(2016)}]{Mikhailov16a}%
  \BibitemOpen
  \bibfield  {author} {\bibinfo {author} {\bibfnamefont {S.~A.}\ \bibnamefont
  {Mikhailov}},\ }\bibfield  {title} {\bibinfo {title} {Quantum theory of the
  third-order nonlinear electrodynamic effects in graphene},\ }\href@noop {}
  {\bibfield  {journal} {\bibinfo  {journal} {Phys. Rev. B}\ }\textbf {\bibinfo
  {volume} {93}},\ \bibinfo {pages} {085403} (\bibinfo {year}
  {2016})}\BibitemShut {NoStop}%
\bibitem [{\citenamefont {Wang}\ \emph {et~al.}(2016)\citenamefont {Wang},
  \citenamefont {Tokman},\ and\ \citenamefont {Belyanin}}]{Wang16}%
  \BibitemOpen
  \bibfield  {author} {\bibinfo {author} {\bibfnamefont {Y.}~\bibnamefont
  {Wang}}, \bibinfo {author} {\bibfnamefont {M.}~\bibnamefont {Tokman}},\ and\
  \bibinfo {author} {\bibfnamefont {A.}~\bibnamefont {Belyanin}},\ }\bibfield
  {title} {\bibinfo {title} {Second-order nonlinear optical response of
  graphene},\ }\href@noop {} {\bibfield  {journal} {\bibinfo  {journal} {Phys.
  Rev. B}\ }\textbf {\bibinfo {volume} {94}},\ \bibinfo {pages} {195442}
  (\bibinfo {year} {2016})}\BibitemShut {NoStop}%
\bibitem [{\citenamefont {Mikhailov}(2017)}]{Mikhailov17a}%
  \BibitemOpen
  \bibfield  {author} {\bibinfo {author} {\bibfnamefont {S.~A.}\ \bibnamefont
  {Mikhailov}},\ }\bibfield  {title} {\bibinfo {title} {Nonperturbative
  quasiclassical theory of the nonlinear electrodynamic response of graphene},\
  }\href@noop {} {\bibfield  {journal} {\bibinfo  {journal} {Phys. Rev. B}\
  }\textbf {\bibinfo {volume} {95}},\ \bibinfo {pages} {085432} (\bibinfo
  {year} {2017})}\BibitemShut {NoStop}%
\bibitem [{\citenamefont {Savostianova}\ and\ \citenamefont
  {Mikhailov}(2017)}]{Savostianova17a}%
  \BibitemOpen
  \bibfield  {author} {\bibinfo {author} {\bibfnamefont {N.~A.}\ \bibnamefont
  {Savostianova}}\ and\ \bibinfo {author} {\bibfnamefont {S.~A.}\ \bibnamefont
  {Mikhailov}},\ }\bibfield  {title} {\bibinfo {title} {Third harmonic
  generation from graphene lying on different substrates: optical-phonon
  resonances and interference effects},\ }\href@noop {} {\bibfield  {journal}
  {\bibinfo  {journal} {Optics Express}\ }\textbf {\bibinfo {volume} {25}},\
  \bibinfo {pages} {3268} (\bibinfo {year} {2017})}\BibitemShut {NoStop}%
\bibitem [{\citenamefont {Cheng}\ \emph {et~al.}(2017)\citenamefont {Cheng},
  \citenamefont {Vermeulen},\ and\ \citenamefont {Sipe}}]{Cheng17}%
  \BibitemOpen
  \bibfield  {author} {\bibinfo {author} {\bibfnamefont {J.~L.}\ \bibnamefont
  {Cheng}}, \bibinfo {author} {\bibfnamefont {N.}~\bibnamefont {Vermeulen}},\
  and\ \bibinfo {author} {\bibfnamefont {J.~E.}\ \bibnamefont {Sipe}},\
  }\bibfield  {title} {\bibinfo {title} {Second order optical nonlinearity of
  graphene due to electric quadrupole and magnetic dipole effects},\
  }\href@noop {} {\bibfield  {journal} {\bibinfo  {journal} {Scientific
  Reports}\ }\textbf {\bibinfo {volume} {7}},\ \bibinfo {pages} {43843}
  (\bibinfo {year} {2017})}\BibitemShut {NoStop}%
\bibitem [{\citenamefont {Alexander}\ \emph {et~al.}(2017)\citenamefont
  {Alexander}, \citenamefont {Savostianova}, \citenamefont {Mikhailov},
  \citenamefont {Kuyken},\ and\ \citenamefont {{Van Thourhout}}}]{Alexander17}%
  \BibitemOpen
  \bibfield  {author} {\bibinfo {author} {\bibfnamefont {K.}~\bibnamefont
  {Alexander}}, \bibinfo {author} {\bibfnamefont {N.~A.}\ \bibnamefont
  {Savostianova}}, \bibinfo {author} {\bibfnamefont {S.~A.}\ \bibnamefont
  {Mikhailov}}, \bibinfo {author} {\bibfnamefont {B.}~\bibnamefont {Kuyken}},\
  and\ \bibinfo {author} {\bibfnamefont {D.}~\bibnamefont {{Van Thourhout}}},\
  }\bibfield  {title} {\bibinfo {title} {Electrically tunable optical
  nonlinearities in graphene-covered {S}i{N} waveguides characterized by
  four-wave mixing},\ }\href@noop {} {\bibfield  {journal} {\bibinfo  {journal}
  {ACS Photonics}\ }\textbf {\bibinfo {volume} {4}},\ \bibinfo {pages} {3039}
  (\bibinfo {year} {2017})}\BibitemShut {NoStop}%
\bibitem [{\citenamefont {Marini}\ \emph {et~al.}(2017)\citenamefont {Marini},
  \citenamefont {Cox},\ and\ \citenamefont {{de Abajo}}}]{MariniAbajo17}%
  \BibitemOpen
  \bibfield  {author} {\bibinfo {author} {\bibfnamefont {A.}~\bibnamefont
  {Marini}}, \bibinfo {author} {\bibfnamefont {J.~D.}\ \bibnamefont {Cox}},\
  and\ \bibinfo {author} {\bibfnamefont {F.~J.~G.}\ \bibnamefont {{de
  Abajo}}},\ }\bibfield  {title} {\bibinfo {title} {Theory of graphene
  saturable absorption},\ }\href@noop {} {\bibfield  {journal} {\bibinfo
  {journal} {Phys. Rev. B}\ }\textbf {\bibinfo {volume} {95}},\ \bibinfo
  {pages} {125408} (\bibinfo {year} {2017})}\BibitemShut {NoStop}%
\bibitem [{\citenamefont {Savostianova}\ and\ \citenamefont
  {Mikhailov}(2018)}]{Savostianova18a}%
  \BibitemOpen
  \bibfield  {author} {\bibinfo {author} {\bibfnamefont {N.~A.}\ \bibnamefont
  {Savostianova}}\ and\ \bibinfo {author} {\bibfnamefont {S.~A.}\ \bibnamefont
  {Mikhailov}},\ }\bibfield  {title} {\bibinfo {title} {Optical {K}err effect
  in graphene: {T}heoretical analysis of the optical heterodyne detection
  technique},\ }\href@noop {} {\bibfield  {journal} {\bibinfo  {journal} {Phys.
  Rev. B}\ }\textbf {\bibinfo {volume} {97}},\ \bibinfo {pages} {165424}
  (\bibinfo {year} {2018})}\BibitemShut {NoStop}%
\bibitem [{\citenamefont {Mikhailov}(2019)}]{Mikhailov19b}%
  \BibitemOpen
  \bibfield  {author} {\bibinfo {author} {\bibfnamefont {S.~A.}\ \bibnamefont
  {Mikhailov}},\ }\bibfield  {title} {\bibinfo {title} {Theory of the strongly
  nonlinear electrodynamic response of graphene: A hot electron model},\
  }\href@noop {} {\bibfield  {journal} {\bibinfo  {journal} {Phys. Rev. B}\
  }\textbf {\bibinfo {volume} {100}},\ \bibinfo {pages} {115416} (\bibinfo
  {year} {2019})}\BibitemShut {NoStop}%
\bibitem [{\citenamefont {Gusynin}\ \emph {et~al.}(2007)\citenamefont
  {Gusynin}, \citenamefont {Sharapov},\ and\ \citenamefont
  {Carbotte}}]{Gusynin07b}%
  \BibitemOpen
  \bibfield  {author} {\bibinfo {author} {\bibfnamefont {V.~P.}\ \bibnamefont
  {Gusynin}}, \bibinfo {author} {\bibfnamefont {S.~G.}\ \bibnamefont
  {Sharapov}},\ and\ \bibinfo {author} {\bibfnamefont {J.~P.}\ \bibnamefont
  {Carbotte}},\ }\bibfield  {title} {\bibinfo {title} {Anomalous absorption
  line in the magneto-optical response of graphene},\ }\href@noop {} {\bibfield
   {journal} {\bibinfo  {journal} {Phys. Rev. Lett.}\ }\textbf {\bibinfo
  {volume} {98}},\ \bibinfo {pages} {157402} (\bibinfo {year}
  {2007})}\BibitemShut {NoStop}%
\bibitem [{\citenamefont {Falkovsky}\ and\ \citenamefont
  {Varlamov}(2007)}]{Falkovsky07a}%
  \BibitemOpen
  \bibfield  {author} {\bibinfo {author} {\bibfnamefont {L.~A.}\ \bibnamefont
  {Falkovsky}}\ and\ \bibinfo {author} {\bibfnamefont {A.~A.}\ \bibnamefont
  {Varlamov}},\ }\bibfield  {title} {\bibinfo {title} {Space-time dispersion of
  graphene conductivity},\ }\href@noop {} {\bibfield  {journal} {\bibinfo
  {journal} {Europ. Phys. J. B}\ }\textbf {\bibinfo {volume} {56}},\ \bibinfo
  {pages} {281} (\bibinfo {year} {2007})}\BibitemShut {NoStop}%
\bibitem [{\citenamefont {Mikhailov}\ and\ \citenamefont
  {Ziegler}(2007)}]{Mikhailov07d}%
  \BibitemOpen
  \bibfield  {author} {\bibinfo {author} {\bibfnamefont {S.~A.}\ \bibnamefont
  {Mikhailov}}\ and\ \bibinfo {author} {\bibfnamefont {K.}~\bibnamefont
  {Ziegler}},\ }\bibfield  {title} {\bibinfo {title} {New electromagnetic mode
  in graphene},\ }\href@noop {} {\bibfield  {journal} {\bibinfo  {journal}
  {Phys. Rev. Lett.}\ }\textbf {\bibinfo {volume} {99}},\ \bibinfo {pages}
  {016803} (\bibinfo {year} {2007})}\BibitemShut {NoStop}%
\bibitem [{\citenamefont {Vasko}\ and\ \citenamefont {Ryzhii}(2008)}]{Vasko08}%
  \BibitemOpen
  \bibfield  {author} {\bibinfo {author} {\bibfnamefont {F.~T.}\ \bibnamefont
  {Vasko}}\ and\ \bibinfo {author} {\bibfnamefont {V.}~\bibnamefont {Ryzhii}},\
  }\bibfield  {title} {\bibinfo {title} {Photoconductivity of intrinsic
  graphene},\ }\href@noop {} {\bibfield  {journal} {\bibinfo  {journal} {Phys.
  Rev. B}\ }\textbf {\bibinfo {volume} {77}},\ \bibinfo {pages} {195433}
  (\bibinfo {year} {2008})}\BibitemShut {NoStop}%
\bibitem [{\citenamefont {Romanets}\ and\ \citenamefont
  {Vasko}(2010)}]{Romanets10}%
  \BibitemOpen
  \bibfield  {author} {\bibinfo {author} {\bibfnamefont {P.~N.}\ \bibnamefont
  {Romanets}}\ and\ \bibinfo {author} {\bibfnamefont {F.~T.}\ \bibnamefont
  {Vasko}},\ }\bibfield  {title} {\bibinfo {title} {Transient response of
  intrinsic graphene under ultrafast interband excitation},\ }\href
  {https://doi.org/10.1103/PhysRevB.81.085421} {\bibfield  {journal} {\bibinfo
  {journal} {Phys. Rev. B}\ }\textbf {\bibinfo {volume} {81}},\ \bibinfo
  {pages} {085421} (\bibinfo {year} {2010})}\BibitemShut {NoStop}%
\bibitem [{\citenamefont {Bao}\ \emph {et~al.}(2010)\citenamefont {Bao},
  \citenamefont {Liu},\ and\ \citenamefont {Lei}}]{Bao10}%
  \BibitemOpen
  \bibfield  {author} {\bibinfo {author} {\bibfnamefont {W.~S.}\ \bibnamefont
  {Bao}}, \bibinfo {author} {\bibfnamefont {S.~Y.}\ \bibnamefont {Liu}},\ and\
  \bibinfo {author} {\bibfnamefont {X.~L.}\ \bibnamefont {Lei}},\ }\bibfield
  {title} {\bibinfo {title} {Hot-electron transport in graphene driven by
  intense terahertz fields},\ }\href@noop {} {\bibfield  {journal} {\bibinfo
  {journal} {Phys. Lett. A}\ }\textbf {\bibinfo {volume} {374}},\ \bibinfo
  {pages} {1266} (\bibinfo {year} {2010})}\BibitemShut {NoStop}%
\bibitem [{\citenamefont {Trushin}\ and\ \citenamefont
  {Schliemann}(2011)}]{Trushin11}%
  \BibitemOpen
  \bibfield  {author} {\bibinfo {author} {\bibfnamefont {M.}~\bibnamefont
  {Trushin}}\ and\ \bibinfo {author} {\bibfnamefont {J.}~\bibnamefont
  {Schliemann}},\ }\bibfield  {title} {\bibinfo {title} {Anisotropic
  photoconductivity in graphene},\ }\href@noop {} {\bibfield  {journal}
  {\bibinfo  {journal} {Europhys. Lett.}\ }\textbf {\bibinfo {volume} {96}},\
  \bibinfo {pages} {37006} (\bibinfo {year} {2011})}\BibitemShut {NoStop}%
\bibitem [{\citenamefont {Shao}\ and\ \citenamefont {Yang}(2015)}]{Shao15}%
  \BibitemOpen
  \bibfield  {author} {\bibinfo {author} {\bibfnamefont {J.~M.}\ \bibnamefont
  {Shao}}\ and\ \bibinfo {author} {\bibfnamefont {G.~W.}\ \bibnamefont
  {Yang}},\ }\bibfield  {title} {\bibinfo {title} {Photoconductivity in {D}irac
  materials},\ }\href@noop {} {\bibfield  {journal} {\bibinfo  {journal} {AIP
  Advances}\ }\textbf {\bibinfo {volume} {5}},\ \bibinfo {pages} {117213}
  (\bibinfo {year} {2015})}\BibitemShut {NoStop}%
\bibitem [{\citenamefont {Singh}\ \emph {et~al.}(2018)\citenamefont {Singh},
  \citenamefont {Ghosh},\ and\ \citenamefont {Agarwal}}]{Singh18}%
  \BibitemOpen
  \bibfield  {author} {\bibinfo {author} {\bibfnamefont {A.}~\bibnamefont
  {Singh}}, \bibinfo {author} {\bibfnamefont {S.}~\bibnamefont {Ghosh}},\ and\
  \bibinfo {author} {\bibfnamefont {A.}~\bibnamefont {Agarwal}},\ }\bibfield
  {title} {\bibinfo {title} {Nonlinear, anisotropic, and giant
  photoconductivity in intrinsic and doped graphene},\ }\href@noop {}
  {\bibfield  {journal} {\bibinfo  {journal} {Phys. Rev. B}\ }\textbf {\bibinfo
  {volume} {97}},\ \bibinfo {pages} {045402} (\bibinfo {year}
  {2018})}\BibitemShut {NoStop}%
\bibitem [{\citenamefont {Ryzhii}\ \emph {et~al.}(2019)\citenamefont {Ryzhii},
  \citenamefont {Ponomarev}, \citenamefont {Ryzhii}, \citenamefont {Mitin},
  \citenamefont {Shur},\ and\ \citenamefont {Otsuji}}]{Ryzhii19}%
  \BibitemOpen
  \bibfield  {author} {\bibinfo {author} {\bibfnamefont {V.}~\bibnamefont
  {Ryzhii}}, \bibinfo {author} {\bibfnamefont {D.~S.}\ \bibnamefont
  {Ponomarev}}, \bibinfo {author} {\bibfnamefont {M.}~\bibnamefont {Ryzhii}},
  \bibinfo {author} {\bibfnamefont {V.}~\bibnamefont {Mitin}}, \bibinfo
  {author} {\bibfnamefont {M.~S.}\ \bibnamefont {Shur}},\ and\ \bibinfo
  {author} {\bibfnamefont {T.}~\bibnamefont {Otsuji}},\ }\bibfield  {title}
  {\bibinfo {title} {Negative and positive terahertz and
  infraredphotoconductivity in uncooled graphene},\ }\href@noop {} {\bibfield
  {journal} {\bibinfo  {journal} {Opt. Mater. Express}\ }\textbf {\bibinfo
  {volume} {9}},\ \bibinfo {pages} {585} (\bibinfo {year} {2019})}\BibitemShut
  {NoStop}%
\bibitem [{\citenamefont {Ignatov}\ and\ \citenamefont
  {Romanov}(1976)}]{Ignatov76}%
  \BibitemOpen
  \bibfield  {author} {\bibinfo {author} {\bibfnamefont {A.~A.}\ \bibnamefont
  {Ignatov}}\ and\ \bibinfo {author} {\bibfnamefont {Y.~A.}\ \bibnamefont
  {Romanov}},\ }\bibfield  {title} {\bibinfo {title} {Nonlinear electromagnetic
  properties of semiconductors with a superlattice},\ }\href@noop {} {\bibfield
   {journal} {\bibinfo  {journal} {phys. stat. sol. (b)}\ }\textbf {\bibinfo
  {volume} {78}},\ \bibinfo {pages} {327} (\bibinfo {year} {1976})}\BibitemShut
  {NoStop}%
\end{thebibliography}

\end{document}